\documentclass[aps,preprint,amsmath,amssymb]{revtex4}
\usepackage{graphicx,color,slashed}
\usepackage{subfigure}
\def\be{\begin{eqnarray}}
\def\ed{\end{eqnarray}}
\def\non{\nonumber}

\begin{document}

\title{ Searching for Vector Dark Matter  via Higgs Portal at the LHC   }

\author{ \bf Chuan-Hung Chen$^{a}$\footnote{Email:
physchen@mail.ncku.edu.tw} and Takaaki Nomura$^{a}$\footnote{Email: nomura@mail.ncku.edu.tw} }

\affiliation{ $^{a}$Department of Physics, National Cheng-Kung
University, Tainan 70101, Taiwan  }

\date{\today}

\begin{abstract}
A Higgs portal dark matter model for explaining the gamma-ray excess from the galactic center can be realized with the extension of local $SU(2)_X$ gauge symmetry with one quadruplet. Due to the residual $Z_3$ discrete symmetry of $SU(2)_X$, the new gauge bosons are the stable dark matter candidates. Due to the mixture of the standard model Higgs doublet and  the introduced quadruplet, dark matter could annihilate into the standard model particles through the Higgs portal and new scalar portal.  We study the discovery significance of the vector dark matter at the Large Hadron Collider. The involved parameters are  consistent with the constraints from relic density and direct detection and with the data of the galactic center gamma-ray excess. With $\sqrt{s}=14$ TeV and luminosities of $100$ and $300$ fb$^{-1}$, we find that a discovery significance of $S/\sqrt{B}=5$ can be easily reached if the production of dark matter is through  the  invisible decays of the Higgs boson and a new scalar boson.

\end{abstract}

\maketitle

\section{Introduction}

Clear evidence of new physics can be based on two measurements, namely neutrino oscillations, which lead to massive neutrinos~\cite{PDG2014}, and astronomical evidence of dark matter (DM), where the weakly interacting massive particles (WIMPs) are the candidates in particle physics. 

Although the direct detection of WIMPs in the LUX experiment~\cite{Akerib:2013tjd} has put a strict limit on the couplings and mass of DM,  potential DM signals are indicated  by the indirect detections, such as  the positron excess, which was uncovered by PAMELA~\cite{Adriani:2008zr} and Fermi-LAT~\cite{FermiLAT:2011ab}. With  measurements of unprecedented precision,  the AMS Collaboration further confirmed the excess  of  the positron fraction in the range of $0.5$ to 500 GeV~\cite{Accardo:2014lma}  and  of  the positron+electron flux from 0.5 GeV to 1 TeV~\cite{Aguilar:2014fea}. 
Nevertheless, there are possible explanations for the cosmic-ray excess, such as pulsars~\cite{pulsar1,pulsar2} and the propagation of cosmic rays from a secondary origin~\cite{Blum:2013zsa}.
 
A clear excess of the gamma-ray spectrum, which has a peak  at the photon energy of around 2 GeV, has recently been found~\cite{Goodenough:2009gk,Hooper:2010mq,Boyarsky:2010dr,Hooper:2011ti,Abazajian:2012pn,Gordon:2013vta,Macias:2013vya,Abazajian:2014fta}. Using the data  of  the {\it Fermi} Gamma-Ray Space Telescope~\cite{Vitale:2009hr, Morselli:2010ty}, a significant signal of gamma rays from the region around the galactic center has been found~\cite{Daylan:2014rsa,Zhou:2014lva,Calore:2014xka,Abazajian:2014hsa,Calore:2014nla}. An interesting finding is that the gamma-ray excess can be interpreted  well  by DM annihilation and  that the associated  thermally averaged cross section $\langle \sigma v_{\rm rel} \rangle$  is on the order of $10^{-26}$ cm$^3$/s, which   is the same as that for the  thermal relic density.  Unlike the case of positron cosmic rays,  a boost factor  is unnecessary for the excess of the gamma-ray spectrum.  Therefore, the gamma-ray spectrum  is a more promising clue for verifying WIMPs as DM candidates. 

With WIMPs considered as DM candidates, it is of interest to determine what effect guarantees the stability of DM and what mediator connects the dark side and visible side. 
To protect DM from decay, a dark charge associated with an unbroken symmetry is  necessary  in the theory;  this charge is usually added to models. Regarding stability,  we assume that an unbroken discrete symmetry can be the remnant of a local gauge symmetry, which is broken spontaneously.   According to previous analysis~\cite{Daylan:2014rsa, Alvares:2012qv, Farzan:2012hh, Alves:2014yha, Berlin:2014pya}, a plausible mechanism for the mediator could be through the Higgs portal~\cite{Patt:2006fw}. The Higgs boson,   the last discovered piece in the standard model (SM) and whose mass is 125 GeV, has been measured recently by ATLAS~\cite{:2012gk} and CMS~\cite{:2012gu}.  In this work, we investigate a model in which  the dark charge is  the residual symmetry of a gauge group  and in which the SM Higgs boson and a new scalar boson are the messengers between dark and visible sectors.

To realize this model, we assume that the DM candidates do not belong to the multiplet of the SM gauge symmetry group,
but are the states of an extra hidden local $SU(2)_X$ gauge symmetry, where $X$ can be regarded as the quantum number of dark charge. Since the local gauge symmetry must accompany  gauge bosons, without further introducing new degrees of freedom, it is plausible to require the new gauge bosons to be the DM candidates. Moreover, in order to have a residual discrete symmetry when the local $SU(2)_X$ is broken spontaneously, we find that our intentions  can be achieved easily if  a quadruplet of $SU(2)_X$ is employed. As a result, a $Z_3$ discrete symmetry  of $SU(2)_X$  remains in the ground state when the quadruplet gets a vacuum expectation value (VEV)~\cite{Chen:2015nea}.  Additionally, the introduced quadruplet is not only  responsible for the breaking of $SU(2)_X$, but also plays an important role in the communication  between dark  and visible sectors. Detailed studies of the model and its implications on relic density and the gamma-ray excess of the galactic center can be found elsewhere~\cite{Chen:2015nea}. Other approaches for stabilizing  the DM in $SU(2)_X$ have been proposed~\cite{Hambye:2008bq,Chiang:2013kqa,Boehm:2014bia,Khoze:2014woa, Baek:2013dwa, Gross:2015cwa, DiChiara:2015bua}.

 Besides direct and indirect DM detection, high-energy colliders, (e.g., Large Hadron Collider (LHC)), could also provide signals of DM.  Such searches depend on the mass of DM and the associated couplings. For $m_{\chi} < m_h/2$, when DM is produced by the on-shell SM Higgs boson, the invisible Higgs decays will directly give a strict constraint on the involved couplings~\cite{Belanger:2013kya, Greljo:2013wja,Aad:2014iia,Chatrchyan:2014tja}. For heavier DM, although the constraint from invisible Higgs decays can be avoided, there is  a  lower production cross section~\cite{Endo:2014cca,Craig:2014lda}. For explaining the gamma-ray excess through DM annihilation, the preferred scale of DM mass is less than the W-gauge boson~\cite{Calore:2014nla,Chen:2015nea}, therefore, we focus on the lighter DM with $m_\chi< m_W$. In this model,  there exists another scalar boson, which is from the quadruplet of $SU(2)_X$. Since the new scalar boson mixes with the SM Higgs, its properties are similar to those of the SM Higgs. We also study its influence on the DM production at colliders.

 Based on the vector DM model, which is dictated by an extra local $SU(2)_X$ gauge symmetry~\cite{Chen:2015nea},  we  study the Higgs-portal vector DM signals at the  $\sqrt{s}=14$ TeV LHC. Besides the background analysis, we discuss 
each channel that produces the DM signal.
 The potential channels include (a) vector boson fusion (VBF), $pp\to S^{(*)} j j$, (b) mono jet, $pp\to S^{(*)} j$, (c) mono W/Z, $pp\to S^{(*)} W/Z$, and (d) $t \bar t$, $pp\to S^{(*)} t \bar t$, where $S^{(*)}$ denotes the on-shell or off-shell Higgs boson and the new scalar boson. We find that the mono jet and VBF channels dominate the DM production cross section, with the other channels having a relatively small contribution. After considering the kinematic cuts for reducing the backgrounds, the ratio of the signal to the background  from the mono jet is smaller than that  from VBF. Therefore, we study the VBF process in detail. 
 
 The rest of this paper is organized as follows:  Section II briefly introduces the WIMP model and summarizes the couplings of DM to the SM Higgs boson and to the new scalar boson.  Section III discusses the constraints of parameters, introduces the signals and possible backgrounds, and analyzes the cross section for each signal channel. We introduce proper kinematic cuts and simulate the signal and background events in Section IV. In addition, we discuss the discovery significance as a function of parameters in this section.  Conclusions are given in Section V. 
  
 \section{  WIMPs in Hidden $SU(2)_X$ and Their Couplings with Higgs}

\subsection { WIMPs}  
 This section briefly introduces the WIMP model and discusses the relevant interactions with DM candidates. 
 For studying the minimal extension of the SM that incorporates DM, besides the SM particles and their associated  gauge symmetry, we consider a new local $SU(2)_X$ gauge symmetry and add one quadruplet of $SU(2)_X$ to the model.  Thus,  the Lagrangian in $SU(2)_X \times SU(2)_L \times U(1)_Y$ is written as:
 \be
 {\cal L}&=& {\cal L}_{SM} + \left( D_\mu \Phi_4\right)^\dagger D^\mu \Phi_4 - V(H,\Phi_4) - \frac{1}{4} X^a_{\mu\nu} X^{a \mu \nu} \label{eq:lang}
 \ed
 with 
 \be
 V(H, \Phi_4) &=& \mu^2 H^\dagger H + \lambda (H^\dagger H)^2 + \mu^2_\Phi \Phi^\dagger_4 \Phi_4 + \lambda_\Phi (\Phi^\dagger_4 \Phi_4)^2 + \lambda'  \Phi^\dagger_4 \Phi_4 H^\dagger H \,, \label{eq:vphi4}
  \ed
where ${\cal L}_{SM}$ is the Lagrangian of the SM, $H^T=(G^+, (v+\phi + i G^0)/\sqrt{2})$ is the SM Higgs doublet, $\Phi^T_4= (\phi_{3/2}, \phi_{1/2}, -\phi_{-1/2}, \phi_{-3/2})/\sqrt{2}$ is the quadruplet of $SU(2)_X$, the index $i$ of $\phi_i$ stands for the eigenvalue of the third generator of $SU(2)_X$, and $\phi_{-i} = \phi^*_{i}$. The covariant derivative of $\Phi_4$ is $D_\mu  = \partial_\mu +i g_X T^a X^a_\mu $ and the representations of $T^3$ is given by $T^3 = {\rm diag}(3/2, 1/2,  -1/2, -3/2)$. The  $T^{1,2}$ can be found elsewhere~\cite{Chen:2015nea}. Since the SM is well known, it is not presented here explicitly.  

For breaking the $SU(2)_X$ and preserving a discrete symmetry,   the non-vanishing VEV and the associated fields that fluctuate around the VEV are set to:
 \be
  \langle \phi_{\pm 3/2} \rangle = \frac{v_4 }{\sqrt{2} }\,, \quad 
  \phi_{\pm 3/2} = \frac{1 }{\sqrt{2} }(v_4 + \phi_r \pm i \xi)\,. \label{eq:vev}
  \ed
  With the breaking pattern  in Eq.~(\ref{eq:vev}), one can find that an $Z_3$ symmetry $U_3 \equiv e^{i  T^3 4\pi/3} = {\rm diag}(1, e^{i2\pi/3}, e^{-2i\pi/3}, 1)$ is preserved by the ground state $\Phi_0=(v_4/2,0,0,v_4/2)$. Under $Z_3$ transformation, the scalar fields of the quadruplet are transformed as:
 \be
\phi_{\pm 3/2} & \longrightarrow&\phi _{\pm 3/2}\,,   \non \\
\phi_{\pm 1/2} & \longrightarrow & e^{\pm i 2\pi/3} \phi_{\pm 1/2}\, . 
  \ed
 That is, $\phi_{\pm 3/2}$ are $Z_3$-blind while $\phi_{\pm 1/2}$ carry the charge of $Z_3$. In terms of the physical states of gauge fields, one can  write: 
 \be
T^a X^a_\mu = \frac{1}{\sqrt{2}}(T^+ \chi_\mu + T^- \bar\chi_\mu ) + T^3 X^3_\mu \label{eq:T+-}
\ed
with $T^\pm = T^1 \pm i T^2$ and $\chi_\mu (\bar \chi_\mu)= (X^1_\mu \mp i X^2_\mu)/\sqrt{2}$, where $\bar\chi_\mu$ is regarded as the antiparticle of $\chi_\mu$. The transformations of gauge fields are  \cite{Chen:2015nea}:
   \be
 &&  X^3_\mu \longrightarrow X^3_\mu \,, \non \\
  && \chi_\mu ( \bar\chi_\mu) \longrightarrow  e^{\mp i 2\pi/3} \chi_\mu (\bar\chi_\mu)\,.
   \ed
It can be seen that  $\phi_{\pm 1/2}$ and  $\chi_\mu (\bar \chi_\mu)$  carry the charges of $Z_3$. Since the masses of $\chi_{\mu}$ and $\bar \chi_\mu$ arise from the spontaneous symmetry breaking of $\Phi_4$. $\phi_{\pm 1/2}$ must be the Nambu-Goldstone bosons  and are the longitudinal degrees of freedom of $\chi_\mu$ and $\bar\chi_\mu$.  Hence, $\chi_\mu$ and $\bar\chi_\mu$ are the candidates of DM in the model. 
  
 \subsection{ Relationships of parameters and couplings to WIMPs} 
 
 In this section, we discuss the new free parameters and their relationships. The new free parameters only appear  in the new gauge sector and  scalar potential shown in Eq.~(\ref{eq:vphi4}). They are  $g_X$, $\mu^2_\Phi$, $\lambda_\Phi$, and $\lambda'$.  In terms of the  SM Higgs doublet and quadruplet of $SU(2)_X$ and  the scalar potential, the  mass matrix for  SM Higgs $\phi$ and new scalar $\phi_r$ is expressed as:
  \be
  M^2   =   \left( \begin{array}{cc}
    m^2_\phi & \lambda' v v_4  \\ 
   \lambda' v v_4 & m^2_{\phi_r} \\ 
  \end{array} \right) \label{eq:Mass}
  \ed
with $m_\phi =\sqrt{2\lambda } v$ and $m_{\phi_r} =\sqrt{2\lambda_\Phi } v_4$. Clearly,  $\lambda'$ causes the mixture of Higgs doublet $H$ and quadruplet $\Phi_4$. The mixing angle  connecting the mass eigenstates  can be parametrized as: 
 \be
  \left( \begin{array}{c}
    h \\ 
    H^0 \\ 
  \end{array} \right) =  \left( \begin{array}{cc}
    \cos\theta & \sin\theta \\ 
    -\sin\theta & \cos\theta \\ 
  \end{array} \right)  \left( \begin{array}{c}
    \phi \\ 
    \phi_r \\ 
  \end{array} \right) \,, \label{eq:mixing}
 \ed
where $h$ denotes the SM-like Higgs boson, $H^0$ is the second scalar boson and  $\tan2\theta = 2\lambda' v v_4 /(m^2_{\phi_r} - m^2_\phi)$. According to Eq.~(\ref{eq:Mass}), the eigenvalues of the mass square matrix are found as:
 \be
 m^2_{h, H^0} = \frac{1}{2} \left(m^2_\phi + m^2_{\phi_r}  \pm \sqrt{(m^2_\phi - m^2_{\phi_r})^2 + 4 \lambda'^2 v^2 v^2_4} \right)\,.
 \ed
We note that the mass of $h$ can be larger or less than that of $H^0$.   In addition,  from the kinetic term of $\Phi_4$, the masses of gauge bosons can be obtained as $m_{\chi} = \sqrt{3}g_X v_4 /2$ and $m_{X^3} = \sqrt{3} m_\chi$. Hence, the set of new free parameters can be chosen as:
 \be
\{ g_X, m_\chi, m_{H^0}, \theta\}\,. 
 \ed
The mixing angle $\theta$ is constrained by the Higgs boson search at the LEP and the LHC. A thorough analysis~\cite{Robens:2015gla} has provided the constraint as a function of the second scalar boson mass. The constraints are taken into account in the analysis of discovery significance below.
 
 Next, we discuss the couplings of DM in the model. Since the DM candidates are the gauge bosons, their couplings to the visible sector are through the mixture of $SU(2)_X$ quadruplet and the SM Higgs doublet.  Therefore, the main interactions of DM are from the kinetic term of $\Phi_4$. With the mixing angle defined in Eq.~(\ref{eq:mixing}), the relevant interactions are given as~\cite{Chen:2015nea}:
  \be
 I_{\chi\bar\chi}& =& \sqrt{3} g_X m_\chi \left( s_\theta h + c_\theta H^0 \right) \chi_\mu \bar\chi^\mu  + \frac{1}{2} \left(\frac{3 g^2_X}{2} \right) \left( s_\theta h + c_\theta H^0 \right)^2 \chi_\mu \bar\chi^\mu  \label{eq:igg}
  \ed
  with $c_\theta=\cos\theta$ and $s_\theta=\sin\theta$.
With these interactions, it can be seen that if $m_h> 2 m_\chi$, then the invisible Higgs decay $h\to \chi \bar\chi$ will give a strict limit on $\sin\theta$. 
Besides the gauge interactions, we also derive  the triple scalar couplings, expressed as:
 \be
 I_{S} = \frac{1}{2} \left(6\lambda v c^2_\theta s_\theta + \lambda' v_4 c^3_\theta \right) h h H^0 
 + \frac{1}{2} \left(-6\lambda_\Phi v_4 c^2\theta s_\theta + \lambda' v c^3_\theta \right) h H^0 H^0\,.
 \ed
Since the mixing angle $\theta$ is related to the parameter $\lambda'$, the two terms in each triple interaction should be the same in terms of their order of magnitude. If the mixing angle is not suppressed,   $H^0$ with $m_{H^0} > m_h/2$ or $m_{H^0} < m_h/2$ through the decay  $H^0\to h h$  or $h\to H^0 H^0$ has an interesting effect on the production of DM. However, since the mixing angle is constrained by DM direct detection and the effects of triple couplings on  the production of  DM pairs are small, we will not further discuss their contributions in this paper.

 \section{ Signals and Backgrounds }
 
 In this section, we  explore the possible DM signals and backgrounds in our model. In order to estimate the cross sections of  signal processes, we firstly discuss various constraints of free parameters and then introduce the possible signals and backgrounds. 
 
 \subsection{ Constraints of free parameters}
 
By Eq.~(\ref{eq:igg}), we see that  the  vector DM candidates only couple to SM Higgs $h$ and new scalar $H^0$. For producing a pair of DM particles, $h$ and $H^0$ could be both off-shell and on-shell. For  the off-shell case, the effect is directly related to the magnitude of couplings and   the main constraints  are from DM relic density and DM direct detection. For the on-shell case, besides the constraints  mentioned above, the invisible Higgs decay  also gives a strong bound.  For presenting the constraint from the invisible Higgs decay, we formulate  the partial decay rate for $h (H^0)\to \chi \chi$ with $m_{h(H^0)} > m_\chi /2$ as:
  \be
  \Gamma( S\to \chi \bar\chi) = \frac{3 g'^2 }{64\pi m_S } \frac{m^4_S - 4 m^2_\chi m^2_S + 12 m^4_\chi}{m^2_\chi} \left( 1- \frac{4 m^2_\chi}{m^2_S}\right)^{1/2}
  \ed
where $g' = g_X s_\theta ( g_X c_\theta)$  when $S=h (H^0)$.  The branching ratios (BRs) of the invisible decays can be expressed as:
\begin{align}
{\rm Br}(h \to \chi \bar \chi) &= \frac{\Gamma_{ h \to \chi \bar \chi}}{\Gamma_{ h \to \chi \bar \chi}+ \Gamma_{h \to {\rm SM}}  c^2_ \theta} 
= \frac{\Gamma_{ h \to \chi \bar \chi}^{g'=1} (g_X \tan \theta)^2 }{\Gamma_{ h \to \chi \bar \chi}^{g' =1} (g_X \tan \theta)^2+ \Gamma_{h \to {\rm SM }}}\,, \\
{\rm Br}(H^0 \to \chi \bar \chi) &= \frac{\Gamma_{ H^0 \to \chi \bar \chi}}{\Gamma_{ H^0 \to \chi \bar \chi}+ \Gamma_{h \to {\rm SM }}^{m_{H^0}} s^2_\theta  }
= \frac{\Gamma_{ H^0 \to \chi \bar \chi}^{g'=1} (g_X \cot \theta)^2 }{\Gamma_{ H^0 \to \chi \bar \chi}^{g'=1} (g_X \cot \theta)^2+ \Gamma^{m_{H^0}}_{h\to {\rm SM }}} \label{eq:inv_H0}
\end{align}
with $\Gamma_{h \to {\rm SM }}$ being the width of the SM Higgs. The expression of $\Gamma_{h \to {\rm SM }}^{m_{H^0}}$ is the same as  that of $\Gamma_{h\to SM}$ but $m_h$ is  replaced by $m_{H^0}$. 

\begin{figure}[hptb] 
\begin{center}
\includegraphics[width=70mm]{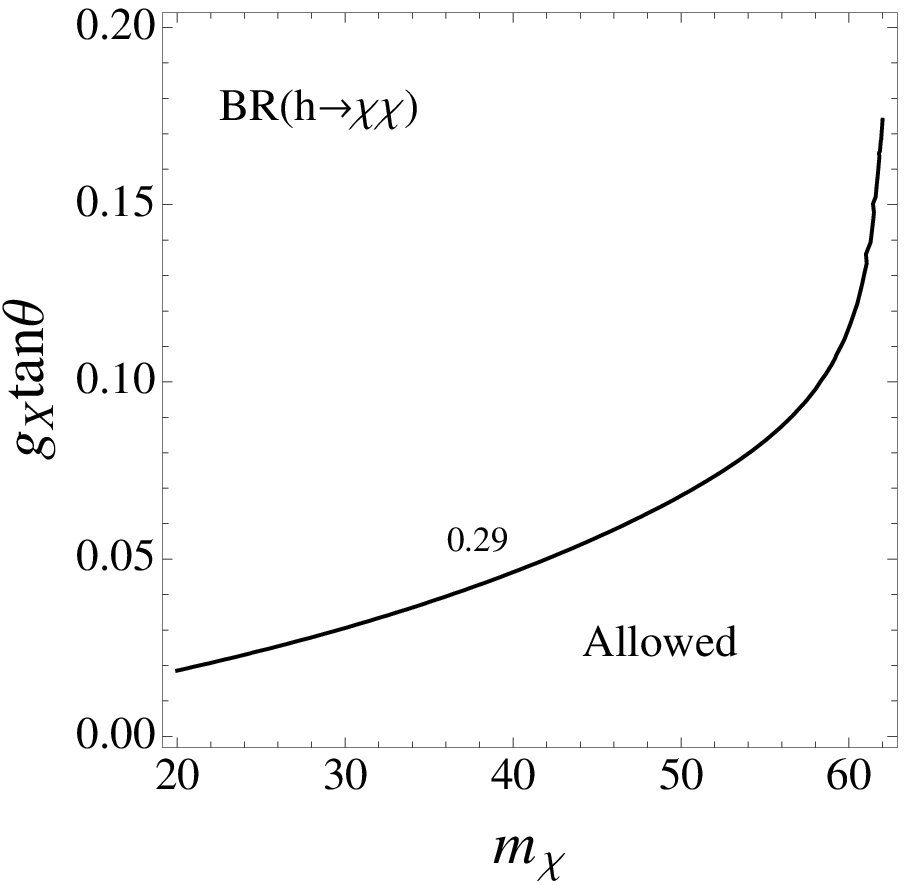} 
\includegraphics[width=70mm]{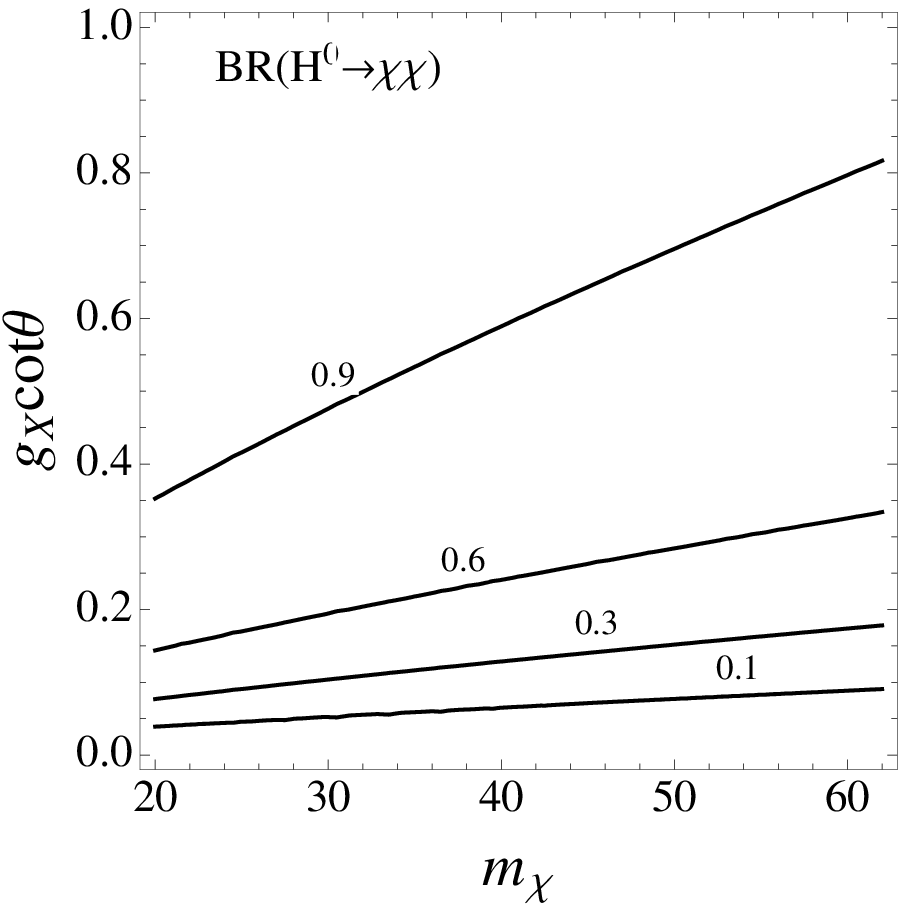} 
\caption{  Contours for invisible decay of SM Higgs $h$ (left) and  of new scalar boson $H^0$ (right) as a function of $g_X \tan\theta(g_X\cot\theta)$ and $m_\chi$, where $m_h=125$ GeV~\cite{:2012gk,:2012gu}, $\Gamma_{h\to SM}=4.21$ MeV~\cite{Heinemeyer:2013tqa}, and the measurement of $BR(h\to \chi \chi)< 0.29$~\cite{Invisible_ATLAS} are used. In $H^0\to \chi \chi$, we assume $m_{H^0}=2 m_\chi +1$.}
\label{fig:Brhxx}
\end{center}
\end{figure}

According to the observations of  ATLAS \cite{:2012gk} and CMS \cite{:2012gu}, the Higgs mass now is known to be $m_h=125$ GeV and the associated width is $\Gamma_{h\to SM}=4.21$ MeV~\cite{Heinemeyer:2013tqa}. Taking these values as inputs, we plot the contours for BR($h\to \chi \chi$) as a function of $g_X \tan\theta$ and $m_\chi$ in left panel of Fig.~\ref{fig:Brhxx}, where the solid line denotes the current  upper limit of data  with $BR(h\to \chi \chi) < 0.29$~\cite{Invisible_ATLAS} and the region above the curve is excluded. Although the new scalar boson $H^0$ has not been observed yet and its mass is unknown, for completeness, we also show its invisible decay as a function of $g_X \cot\theta$ and $m_\chi$ in the right panel of Fig.~\ref{fig:Brhxx} with the setting of  $m_{H^0}=2 m_\chi +1$ GeV, where  the adopted mass relation $m_{H^0} \simeq 2 m_\chi$  can  explain the galactic center gamma-ray excess~\cite{Chen:2015nea}. 

\begin{figure}[hptb]
\begin{center}
\includegraphics[width=80mm]{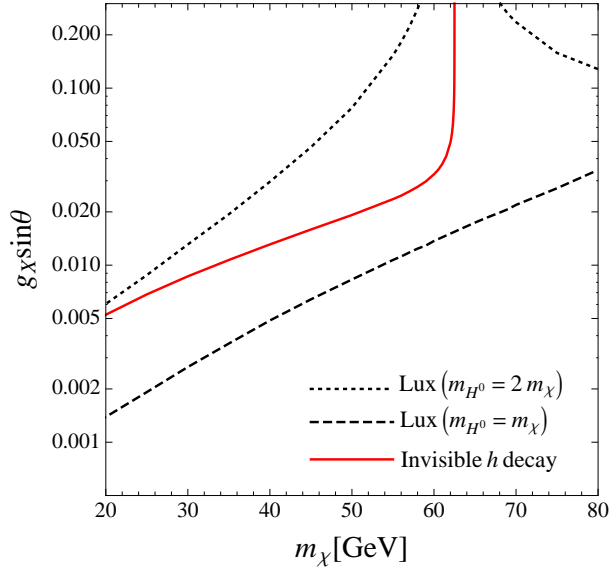}
\end{center}
\caption{ Bounds of $g_X \sin\theta $ and $m_\chi$ from DM scattering off nucleons, where the red solid and black dashed (dotted) lines stand for  the limits from invisible SM Higgs decay and LUX experiment ~\cite{Akerib:2013tjd} with $m_{H^0} = m_\chi ( 2 m_\chi)$, respectively. }
\label{fig:Limit}
\end{figure}

Based on a previous investigation~\cite{Chen:2015nea}, it is known that  the measured relic density of DM could bound the couplings of DM annihilating into SM particles; however, a stronger limit has arisen from the direct detection. By Eq.~(\ref{eq:igg}), we see that the coupling of each $h$ and $H^0$ to $\chi$ is associated with $s_\theta$ and $c_\theta$, respectively. 
Since their couplings to quarks are $c_\theta$ and $s_\theta$, except the mass differences of intermedia, the spin-independent cross section of DM scattering  off nucleons only depends on $g_X c_\theta s_\theta$ for both  $h$ and $H^0$.  
 In terms of the results measured by the LUX Collaboration~\cite{Akerib:2013tjd},  we present the allowed values of $g_X s_\theta$ and $m_X$ in Fig.~\ref{fig:Limit}, where the dashed and dotted lines stand for $m_{H^0}=m_\chi$ and  $2 m_\chi$, respectively.  For comparison, we also show the situation for invisible Higgs decay $h\to \chi \bar\chi$. From the results, it can be seen that for $m_\chi < m_{h(H^0)}/2$, invisible Higgs decay gives the most restrictive limit. However, for $m_{H^0} \sim m_\chi$, the strongest bound is from the experiment of DM direct detection.   

\subsection{ Signal processes and backgrounds}

The signals of WIMPs in the model originate from the SM Higgs and new scalar $H^0$ decays, denoted by $S^{(*)}\to \chi \bar\chi$, in which scalar $S$ can be on-shell or off-shell. The  potential channels for producing DM pairs through the $S$-portal are found to be (a) VBF, $pp\to S^{(*)} j j$, (b) mono jet, $pp\to S^{(*)} j$, (c) mono W/Z, $pp\to S^{(*)} W/Z$, and (d) $t \bar t$, $pp\to S^{(*)} t \bar t$. The mono jet is the loop-induced process $gg \to S^{(*)} g$~\cite{Goodman:2010ku,Fox:2011pm}. In order to  understand and estimate the production cross section of each channel, we implemented our model in CalcHEP~\cite{Belyaev:2012qa} and utilized the code with {\tt CTEQ6L} PDF~\cite{Nadolsky:2008zw}  and $\sqrt{s}=14$ TeV to run numerical calculations.  

\begin{figure}[phtb]
\begin{center}
\includegraphics[width=70mm]{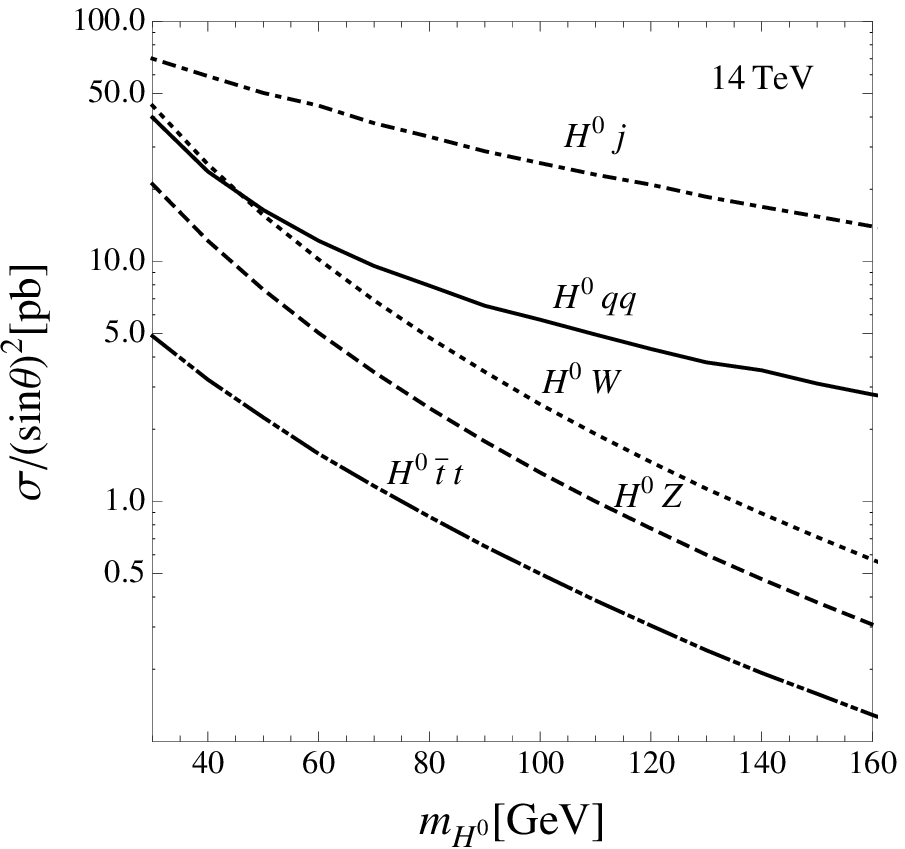}
\includegraphics[width=70mm]{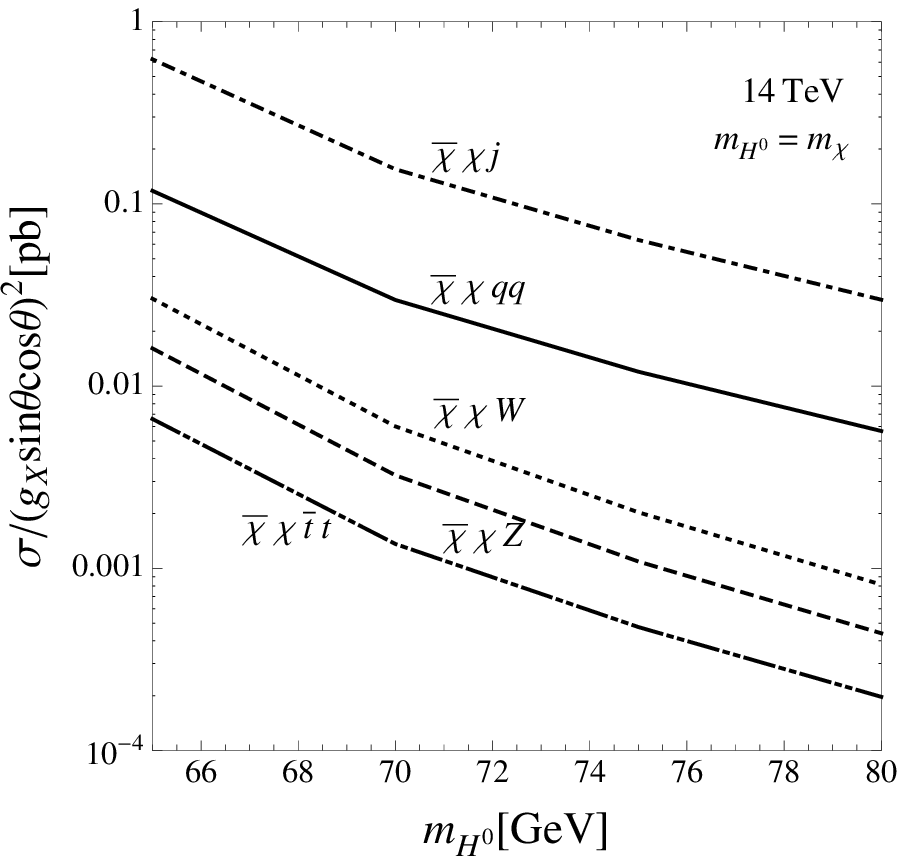}
\end{center}
\caption{Cross section of each signal channel for on-shell $H^0$ (left) and off-shell $h$ and $H^0$ (right) as a function of $m_{H^0}$, where $\sqrt{s}=14$ TeV is used and for each panel,  we scaled the cross section by factors of $1/\sin^2\theta$ and $1/(g^2_X \sin^2\theta \cos^2\theta)$, respectively.   }
\label{fig:S}
\end{figure}

Consequently, the production cross sections for $pp\to S^{(*)} X$, with X being the involved final state, are presented  in Fig.~\ref{fig:S}, where the left  panel is for on-shell $H^0$. Since the effect of on-shell $h$  is similar to that of $H^0$, except for  $c_\theta$ dependence  instead of $s_\theta$, here we only show the results of $H^0$. The right panel of Fig.~\ref{fig:S} is for off-shell $h$ and $H^0$. In this case, since  $m_\chi$ and $m_{H^0}$ are the explicit parameters in the processes, we adopt $m_{H^0}=m_{\chi}$ as the representative case. For reducing the dependence of this parameter, we scale the left and right panels by  factors of $1/s^2_\theta$ and $1/(g^2_X s^2_\theta c^2_\theta)$, respectively. 
 From these results,  it can be clearly seen that mono jet and VBF processes dominate the production cross section in the region of $m_{H^0} > 50$ GeV. Nevertheless, when we further impose the kinematic cuts for reducing the events of backgrounds, the contributions of mono jet to the significance will become sub-leading effects. Hence, the main contributions to the signals  are indeed from the VBF process and thus we focus our study on this channel. Additionally, from the plots, we also know that the off-shell processes are much smaller than those arising from the resonance of $S$. 

Since DM candidates are invisible particles and the production mechanism at the LHC in the model is through  VBF, the signal events at the  detector level will appear as:
\begin{equation}
\label{eq:signal}
{\rm 2 \ jets +   \slashed{E}_T}\,,
\end{equation}
where $\slashed{E}_T$ is the missing transverse energy.
As known, the background events generated  from the SM contributions can also mimic the signal events of Eq.~(\ref{eq:signal}).
In order to distinguish  the signals from the backgrounds, we consider the following background processes~\cite{Ghosh:2012ep}: \\
(1) $Z jj$ background : $p p \to Z j j$\,, \\
(2) $Z jjj$ background : $p p \to Z j j j$\,, \\
(3) $W j j $ background : $p p \to W^\pm j j$, \\
(4) $W j j j$ background : $p p \to W^\pm j j j$, \\
(5) top background : $p p \to t W^- \bar b (\bar t W^+ b)$, \\
where the missing transverse energy $\slashed{E}_T$ is from the $Z$- and $W$-boson  leptonic decays. Although charged leptons can be generated by $W$ decays,  when they are misidentified by the detectors, the events will appear as  missing $E_T$.  Similarly, this situation could also happen in jet. Therefore, for analyzing the backgrounds, when the events are generated, we set the number of jets in the final states to be up to three. 
We note that although QCD multi jets  are also the source of background,  however their contributions can be further reduced by the kinematic cuts.

\section{Event simulation and discussions}

After discussing the potential DM signals and possible backgrounds, we simulate the events by introducing  proper kinematic cuts and investigate the resultant significance at $\sqrt{s}=14$ TeV and luminosities of 100 and 300 fb$^{-1}$. As mentioned earlier, since  the VBF process $p p \to S^{(*)} q q $ is  the most promising mode to get a large ratio of the signal to background, in the following analysis, we only concentrate on VBF. 

In order to perform the event simulation, we employ the event generator {\tt MADGRAPH/MADEVENT\,5} (MG5)~\cite{Ref:MG} with  {\tt NNPDF23LO1} PDFs~\cite{Deans:2013mha}, where the necessary Feynman rules and relevant parameters of the model are created by {\tt FeynRules 2.0} \cite{Alloul:2013bka}. The generated events are further passed onto {\tt PYTHIA\,6}~\cite{Ref:Pythia} to deal with the fragmentation of hadronic effects,  the  initial-state radiation (ISR) and final-state radiation (FSR) effects, and the decays of SM particles (e.g. $Z$-boson, $W$-boson, $t$-quark, etc.). In addition, these events are also run through the {\tt PGS\,4} detector simulation~\cite{Ref:PGS}.

\subsection{ Event selections and kinematic cuts}

\begin{figure}[hptb]
\begin{center}
\includegraphics[width=70mm]{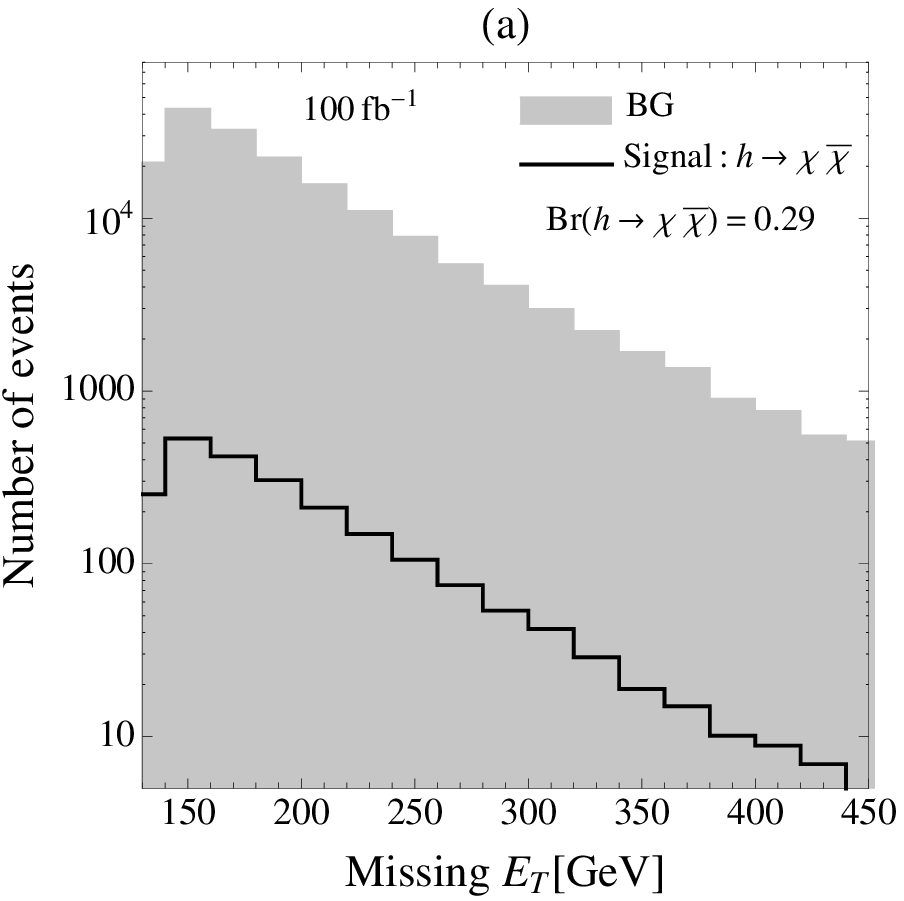}
\includegraphics[width=70mm]{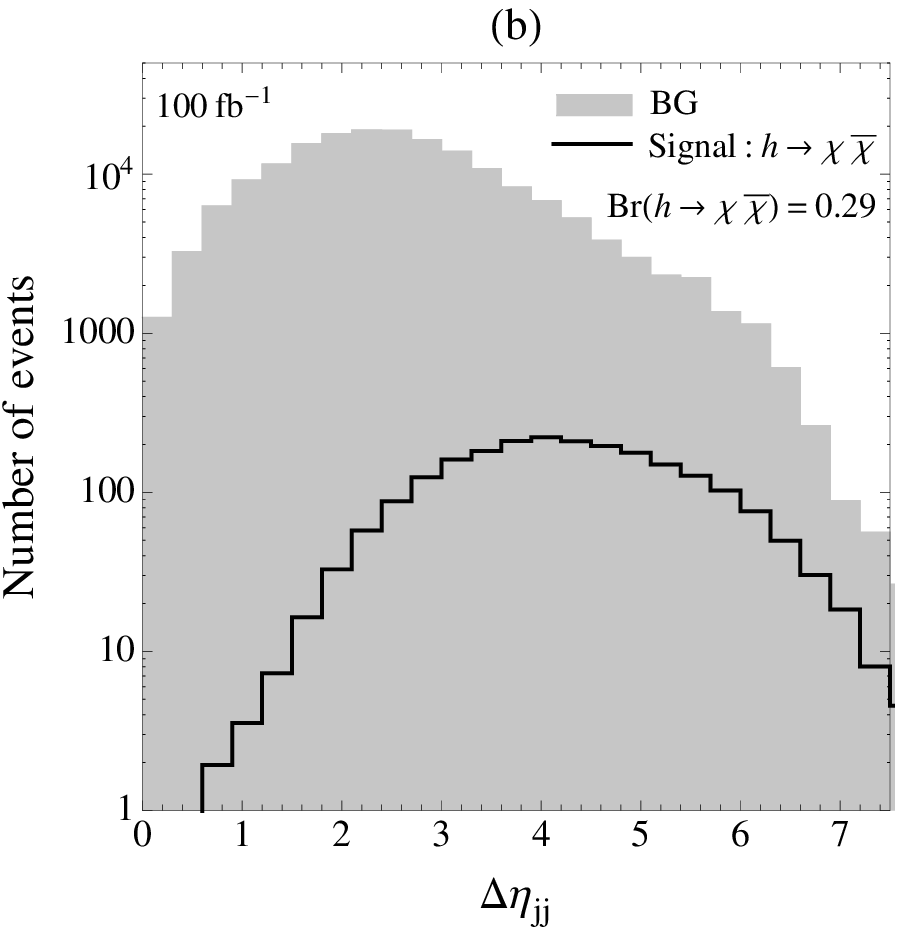}
\includegraphics[width=70mm]{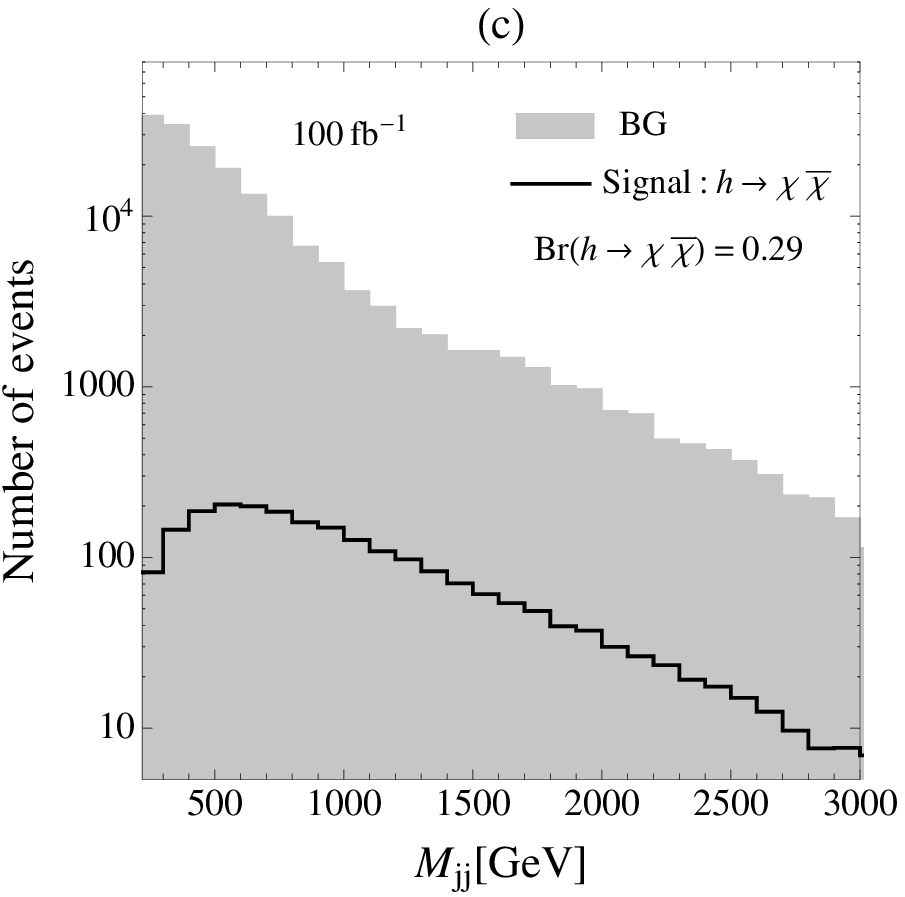}
\includegraphics[width=70mm]{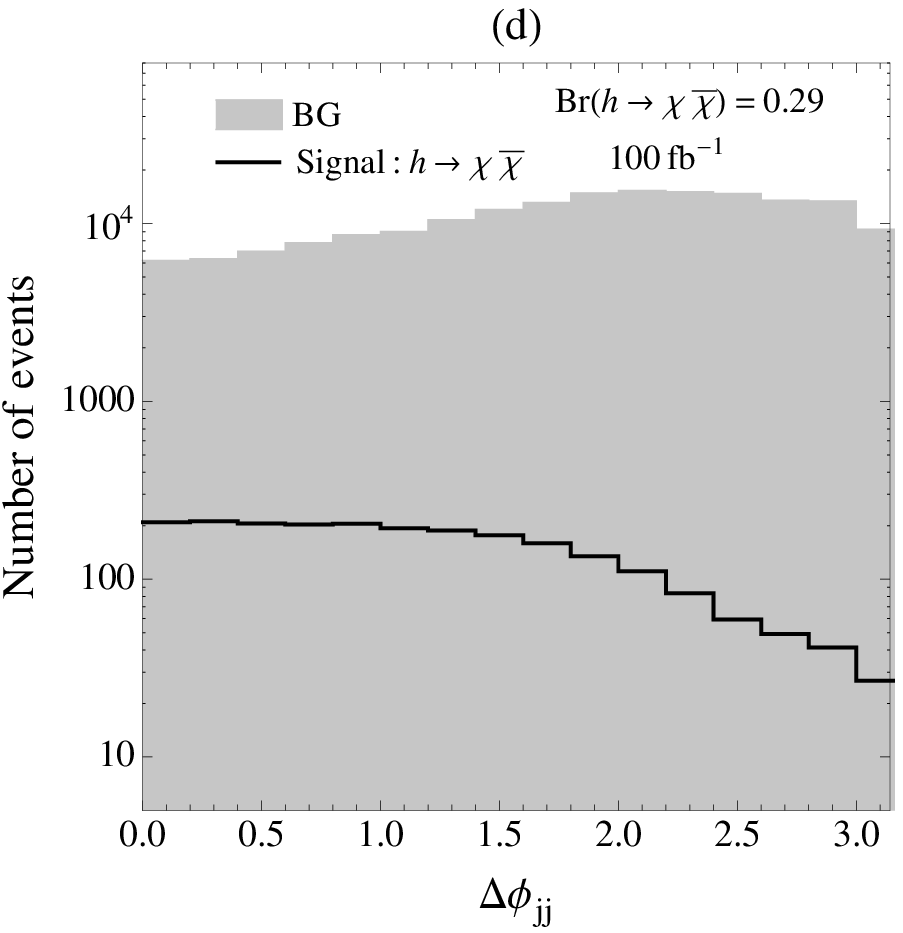}
\end{center}
\caption{Event histogram for $pp\to h(\chi \bar\chi) qq$ and related background as a function  of (a) missing $E_T$, (b)$\Delta \eta_{jj}$, (c) $M_{jj}$, and (d) $\Delta \phi_{jj}$, where the event selection criteria of   Eq.~(\ref{eq:Cut1}) are adopted. }
\label{fig:distBC}
\end{figure}

For enhancing the ratio of the signal to background,  we propose  proper criteria to suppress the backgrounds. Since Higgs portal DM production at the LHC has been studied in the literature~\cite{Djouadi:2011aa, Djouadi:2012zc} and the production mechanism also exists in our model,  we first  perform the DM production through the processes  $pp\to h qq$ and  invisible Higgs decay $h\to \chi \bar\chi$. Then,  we can directly apply  the event selections  for invisible Higgs search proposed by CMS~\cite{Chatrchyan:2014tja} and set them as
\be
 p_T(j) > 50 \ {\rm GeV}, \, \, |\eta(j)| < 4.7, \, \, \eta(j_1) \cdot \eta(j_2) < 0, \, \, \slashed{E}_{T} > 130 \ {\rm GeV}, \, \, M_{jj} > 200 \ {\rm GeV} \,,\label{eq:Cut1}
\ed
where $p_T(j)$ and  $\eta(j)$ are the transverse momentum and pseudo-rapidity of jet $j$, and $M_{jj}$ is the invariant mass of two jets. 
 Although these conditions are used for collision energy at $\sqrt{s}= 7$ and 8 TeV in the CMS experiments, according to our MG5 event  simulations,  it is found that the distributions of jet $p_T$ and $\slashed{E}_T$  indeed are not sensitive to the total collision energy of LHC.  Therefore, in this study we take these selection conditions as the basic criteria for event kinematic cuts at $\sqrt{s}=14$ TeV.  
Using the event generator MG5 and the cuts of Eq.~(\ref{eq:Cut1}), we show the histograms of the signal and background as a function of ${\slashed E}_T$, $\Delta \eta_{jj}$, $M_{jj}$ and $\Delta \phi_{jj}$ in Figs.~\ref{fig:distBC}(a), (b), (c), and (d), respectively, where  $\Delta \eta_{jj}= | \eta_{j_1} - \eta_{j_2} |$ and $\Delta \phi_{jj}= |\phi_{j_1} - \phi_{j_2} | $ are the pseudo-rapidity difference and azimuthal angle difference of the two-jet final state, respectively. 

\begin{figure}[hptb]
\begin{center}
\includegraphics[width=70mm]{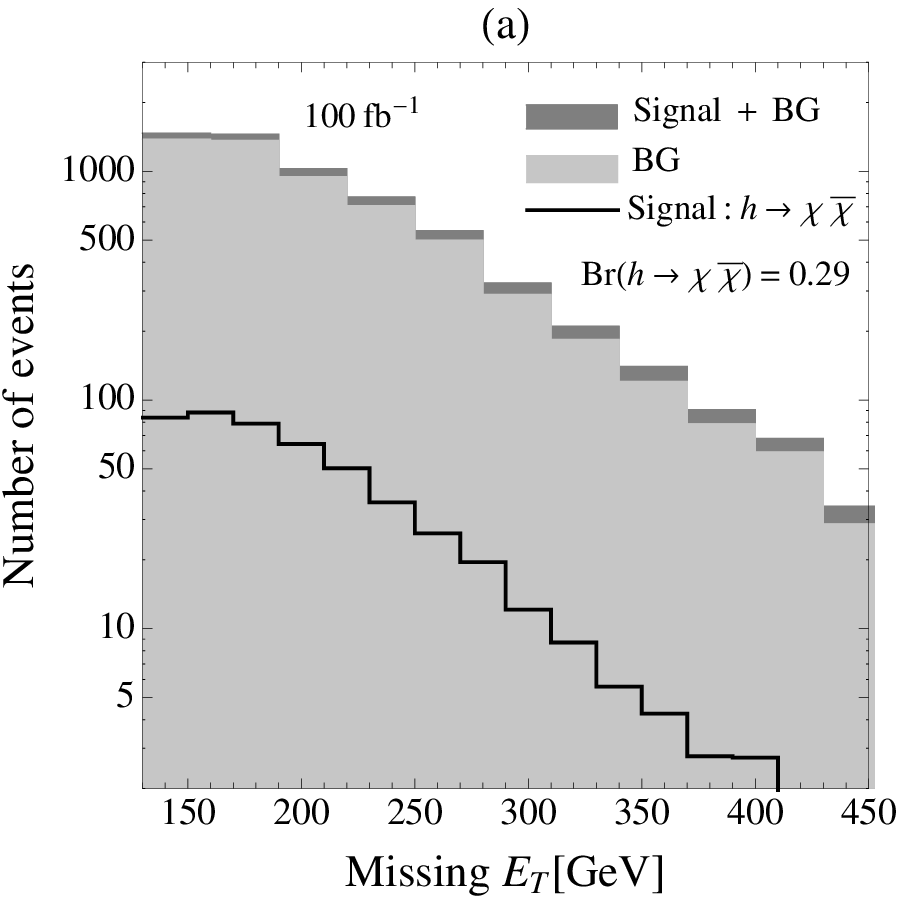}
\includegraphics[width=70mm]{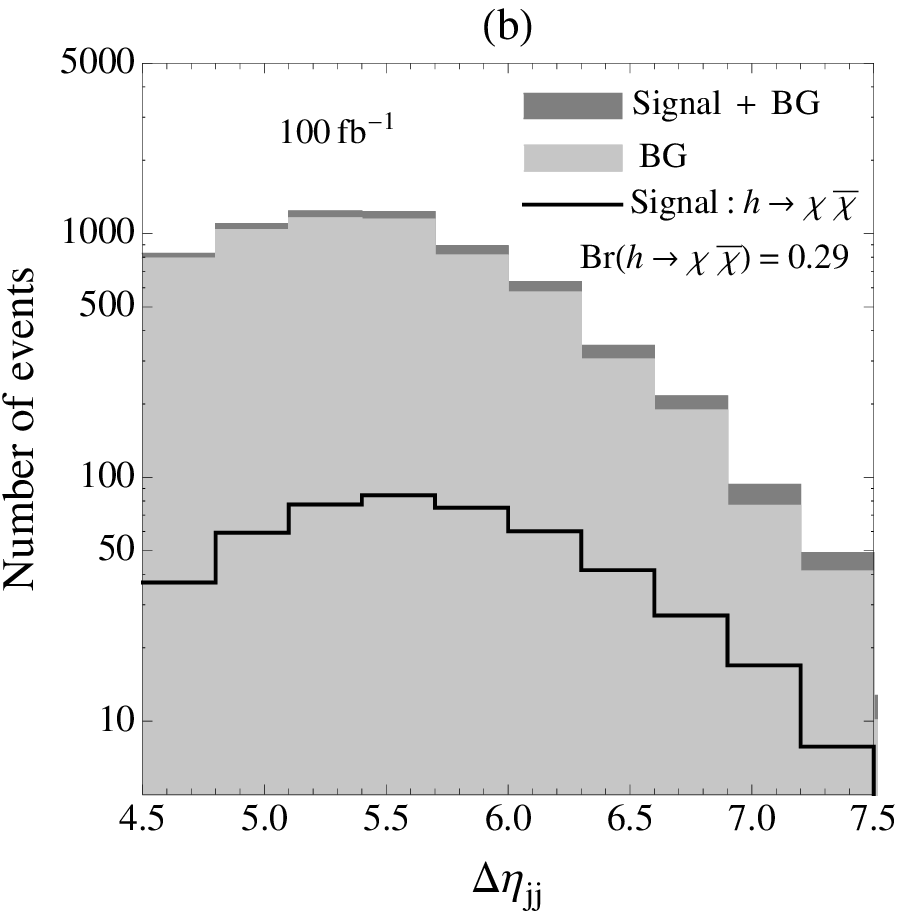}
\includegraphics[width=70mm]{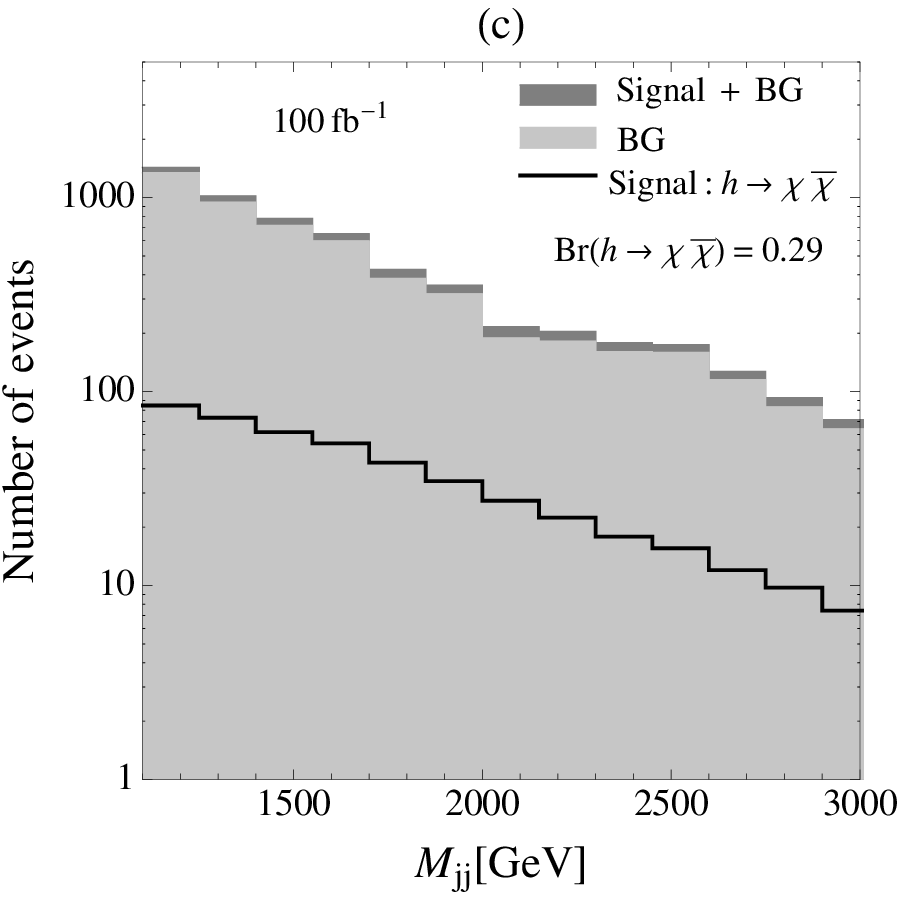}
\includegraphics[width=70mm]{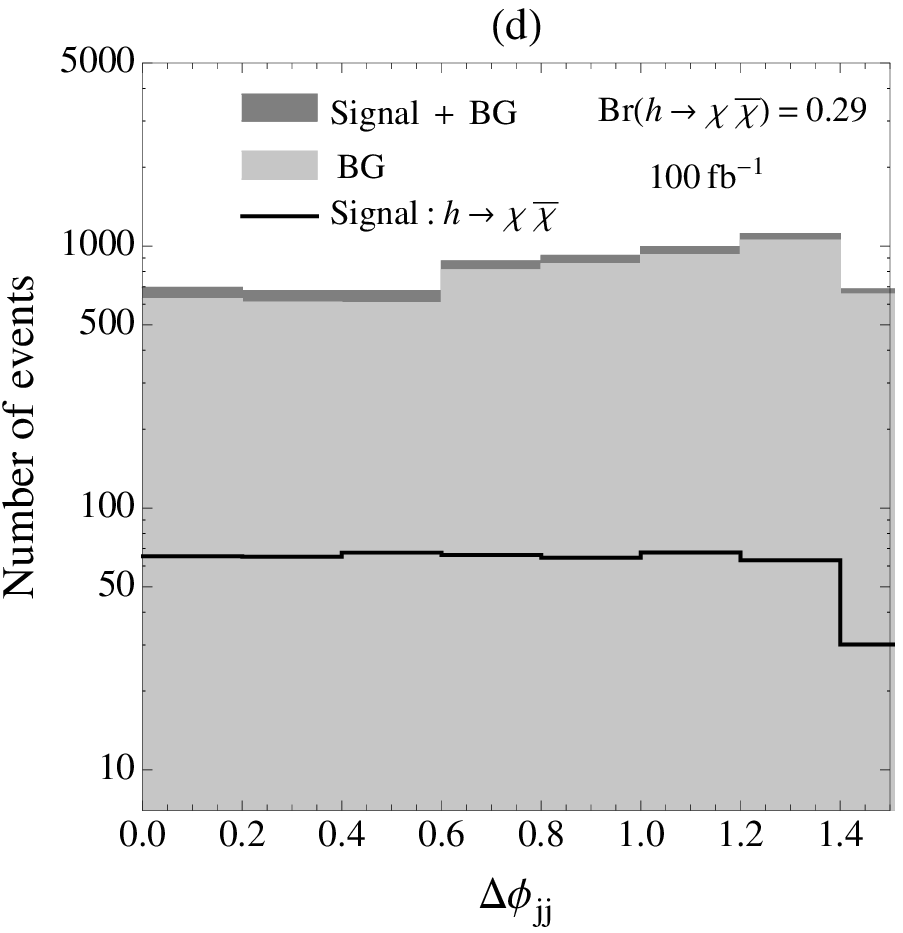}
\end{center}
\caption{The legend is the same as that in Fig.~\ref{fig:distBC}, but both basic and advanced cuts of Eqs.~(\ref{eq:Cut1}) and (\ref{eq:Cut2}) are applied simultaneously.} 
\label{fig:distAC}
\end{figure}

 From the results of Fig.~\ref{fig:distBC}, we find that  at low $\Delta\eta_{jj}$, background events are much larger than signals. If we further impose a cut on $\Delta\eta_{jj}$, the background events will be significantly reduced. Similar behavior also occurs  at $M_{jj}< 1100$ GeV and $\Delta\phi_{jj}> 1.5$. Therefore, utilizing the difference  in  the kinematic region between the signal and background, we propose stricter event selection conditions on $\Delta\eta_{jj}$, $M_{jj}$, and $\Delta\phi_{jj}$,  given as:
\be
 \quad \Delta \eta_{jj} > 4.5, \quad \Delta \phi_{jj} < 1.5, \quad M_{jj} > 1100 \ {\rm GeV}\,.\label{eq:Cut2}
\ed
When both cuts of Eqs.~(\ref{eq:Cut1}) and (\ref{eq:Cut2}) are imposed simultaneously,  the resultant histograms as a function of ${\slashed E}_T$, $\Delta \eta_{jj}$, $M_{jj}$, and $\Delta \phi_{jj}$ are shown in Fig.~\ref{fig:distAC}. 

\subsection{Discovery significance of signal}

The event production of $pp\to S^{(*)} (\to \chi \bar\chi) X$ depends on the mass of DM. For interpreting the gamma-ray excess from the galactic center at the same time,  we concentrate on the DM with $m_\chi < m_W$. For distinguishing the contributions of on-shell $h$ from the off-shell one, we set the allowed range of $m_\chi$ to be the following two schemes:
\begin{align}
\label{eq:Mh}
 {\rm I}: \quad m_\chi < \frac{m_h}{2} \,, \qquad {\rm II}: \quad \frac{m_{h}}{2} <  m_{\chi} < m_W\,,
\end{align}
where the former dictates DM pair production to be through invisible Higgs decay while the latter dictates it to be through the Higgs propagator. Based on a previous study~\cite{Chen:2015nea}, for explaining the galactic center gamma-ray excess  via DM annihilation in this model, the favored ranges for $m_{H^0}$ are $m_{H^0} \gtrsim 2 m_\chi$ and $m_{H^0} \lesssim  m_{\chi}$. In order to fit well with gamma-ray data, we find $m_{H^0} \sim 2m_\chi$ and $m_\chi \sim m_{H^0}$. 
For numerical calculations, we adopt the mass relation as: 
\begin{align}
\label{eq:MH}
 {\rm A}: \quad m_{H^0} > 2 m_\chi, \qquad
{\rm B}: \quad m_{H^0} <  2 m_{\chi}  \ {\rm GeV}\,.
\end{align}
Here, $H^0$ can decay into a DM pair in case ${\rm A}$ while case ${\rm B}$ is through off-shell $H^0$. Since $m_h=125$ GeV is known and the unobserved $m_{H^0}$ is still a free parameter, we investigate various situations by combining schemes ${\rm I}$ and ${\rm II}$ of Eq.~(\ref{eq:Mh}) with cases $A$ and $B$ of Eq.~(\ref{eq:MH}), denoted as $\rm I_A$, $\rm I_B$, $\rm II_A$, and $\rm II_B$,

After setting up the kinematic cuts and classifying the possible schemes for $m_\chi$ and $m_{H^0}$, we calculate the cross sections of the background and signal with various values of $m_\chi$ in schemes $\rm I_{A,B}$ and $\rm II_{A,B}$. The numerical values are presented  in Table~\ref{tab:Nevent}, where the simulated events had passed through ${\tt PYTHIA\,6}$ and ${\tt PGS\,4}$ detector simulation, the cuts of Eqs.~(\ref{eq:Cut1}) and (\ref{eq:Cut2}) were employed, $R_{h} = c^2_\theta BR(h \to \chi \bar \chi) $, $R_{H^0} = s^2_\theta BR(H^0 \to \chi \bar \chi)$, and $R_{\rm off} = (g_X s_\theta c_\theta)^2$. The associated cross section is obtained as:
 \be
 \sigma_{BG/S} = \sigma^{\rm MG5}_{BG/S} \frac{N_{\rm cuts}}{N_{\rm tot}}\,.
 \ed
Here, $\sigma^{\rm MG5}_{BG/S}$ is the cross section of the background/signal provided by MG5, $N_{\rm tot}$ is the number of original generated events and $N_{\rm cuts}$ is the number of selected events.  
For the background events, we have summed up all channels. We find that  the dominant backgrounds are from $Z$ + jets,  where $Z$ invisibly decays into neutrinos. By requiring null charged leptons in the final states, the event number from $W$ + jets should be smaller than that from $Z$ + jets. 
We also investigate the background associated with the top-quark by generating event $t W^- \bar b (\bar t W^+ b)$, which includes $t \bar t$ production. Since  the corresponding cross section  is less than 1 fb when event selections are applied,  we do not show its value in Table.~\ref{tab:Nevent}. 
 To understand the effect of kinematical cuts, we show the background cross section  for each step of cuts,  where the basic cuts are shown in  Eq.~(\ref{eq:Cut1}). It can be clearly seen that $\Delta \eta_{jj}$ cuts reduce the background significantly.
   \begin{table}[hptb]
   \caption{Cross sections of signal and background (in units of fb) at $\sqrt{s}=14$ TeV and at detector level, where the introduced kinematic cuts in Eqs.~(\ref{eq:Cut1}) and (\ref{eq:Cut2}) were applied,  $R_{h} = \cos^2\theta BR(h \to \chi \bar \chi) $, $R_{H^0} = \sin^2 \theta BR(H^0 \to \chi \bar \chi)$, and 
   $R_{\rm off} = (g_X \sin \theta \cos \theta)^2$. For the background, we present the cross sections after each cut, where the basic cuts are shown in Eq.~(\ref{eq:Cut1}). }
  \label{tab:Nevent}
  \begin{tabular}{c||ccccc}  \hline
   &  &  Zjj & Zjjj & Wjj & Wjjj \\ \cline{2-6}
   & Basic cuts &  2831. & 705. & 1315. & 184. \\
 $\sigma_{\rm BG}$[fb]  & + $\Delta \eta_{jj}$ & 124. & 33.8  & 50.6  & 7.73 \\
   & + $\Delta \phi_{jj}$ & 69.4 & 18.1 & 26.2 & 3.97 \\
   & + $M_{jj}$ &  32.9 & 8.54 & 14.2 & 2.20  \\ \hline \hline
   & $m_\chi$[GeV]  & 40 & 50 & 60 &  \\
   & I$_{\rm A}$  & 18.6 $R_{H^0}$ + 17.2 $R_h$  \qquad   & \qquad 17.5 $R_{H^0}$ + 17.2 $R_h$  \qquad   &  \qquad 17.3 $R_{H^0}$ + 17.2 $R_h$ &  \\
 $\sigma_{\rm S}$[fb]  & I$_{\rm B}$  & 17.2 $R_h$ & 17.2 $R_h$ & 17.2 $R_h$ & \\ \cline{2-6}
   & $m_\chi$[GeV]  &  65   &  70   & 75 & \\
   & II$_{\rm A}$  &   16.3 $R_{H^0}$  &  16.0 $R_{H^0}$     &   15.4 $R_{H^0}$  &        \\
   & II$_{\rm B}$  &  0.689 $R_{\rm off}$ & 0.211 $R_{\rm off}  $     &  0.102 $R_{\rm off} $ & \\ \hline
  \end{tabular}
\end{table}

\begin{figure}[hptb]
\begin{center}
\includegraphics[width=70mm]{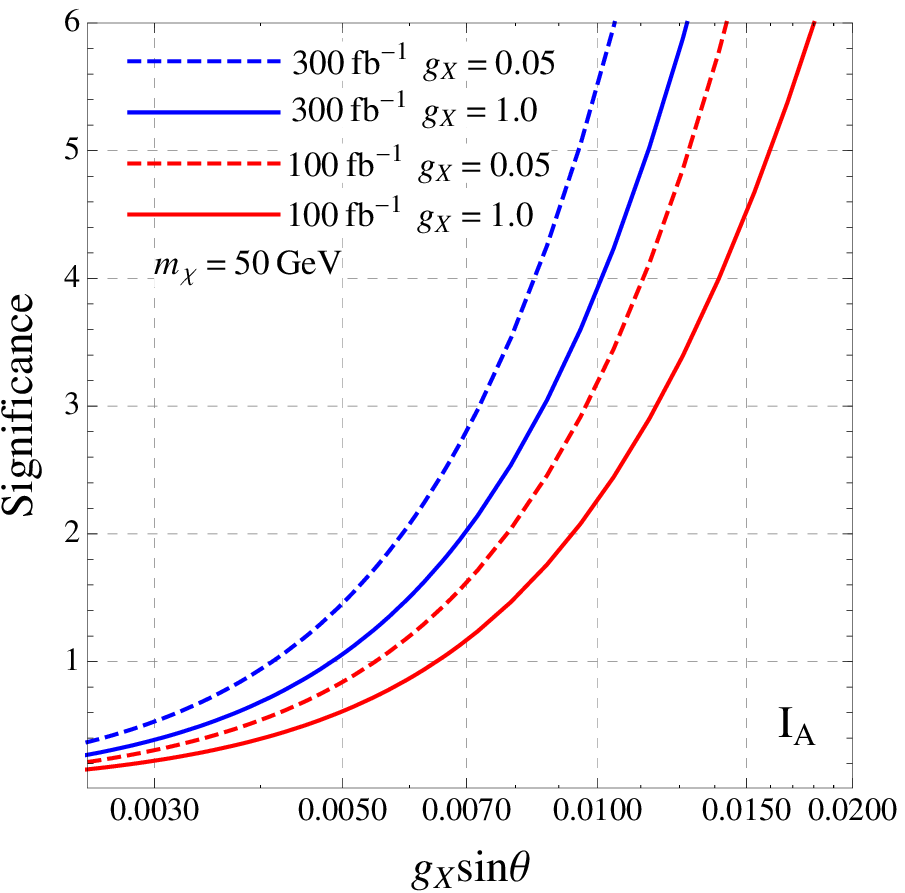}
\includegraphics[width=70mm]{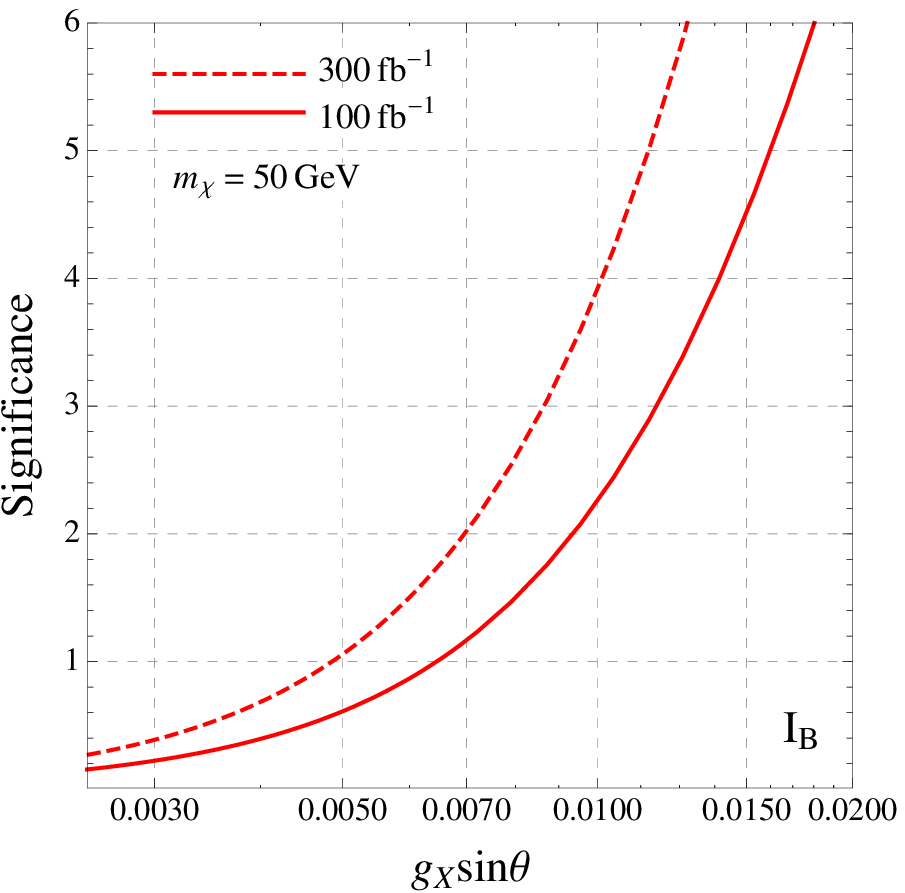}
\includegraphics[width=70mm]{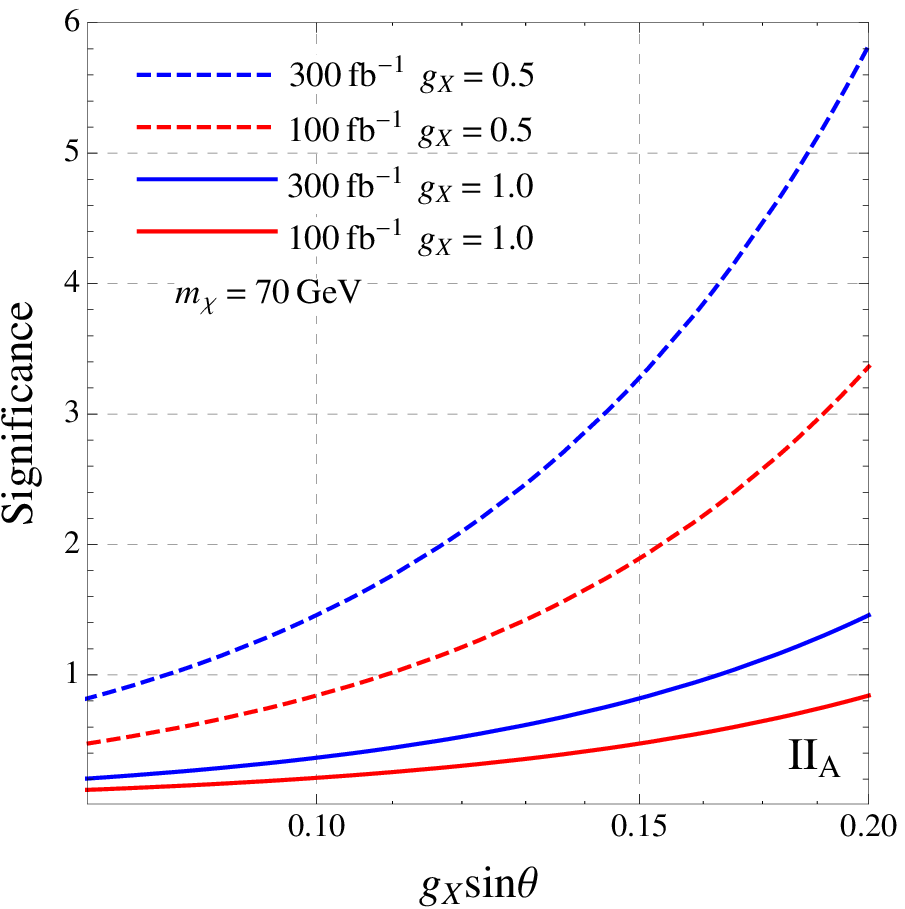}
\includegraphics[width=70mm]{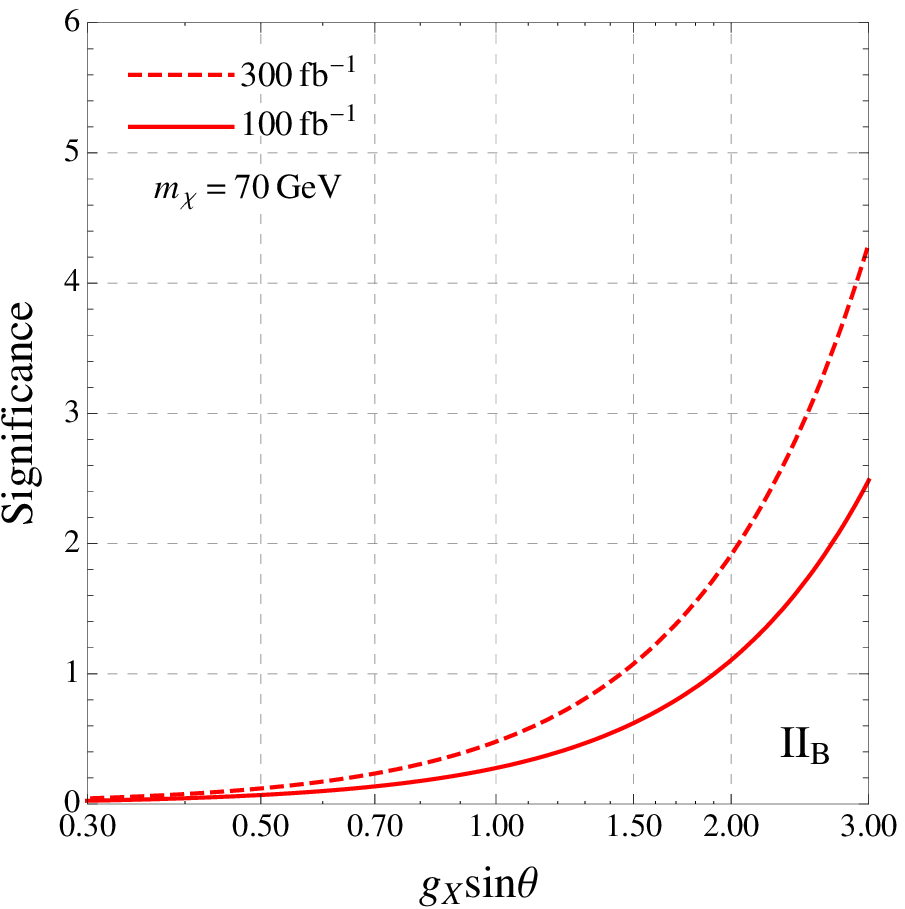}
\end{center}
\caption{Discovery significance as a function of $g_X \sin\theta$, where we set $m_\chi =50$ GeV for ${\rm I_{A,B}}$ and $m_\chi=70$ GeV for ${\rm II_{A,B}}$  and the luminosities of 100 and 300 fb$^{-1}$ are used. }
\label{fig:Sig_Coupling}
\end{figure}

For studying the potentiality  of discovery,  as typically done, we define the significance as  $S=n_s/\sqrt{n_b}$, 
where $n_s$ and $n_b$ denote the numbers of selected events for the signal and background, respectively. 
For numerical illustration, we take $m_\chi=50$ GeV for the schemes $\rm I_{A, B}$  and  $m_\chi =70$ GeV for schemes ${\rm II_{A,B}}$. 
Here, we adopt $m_H = m_\chi -1$ GeV and $m_H = 2 m_\chi + 1$ GeV for schemes ${\rm \{ I_B, II_B \} }$ and ${\rm \{ I_A, II_A \} }$, respectively.
Accordingly, we display the discovery significance as a function of $g_X s_\theta$ with 100 and 300 fb$^{-1}$ in Fig.~\ref{fig:Sig_Coupling}. Since the sensitive regions of $g_X s_\theta$ are different in different schemes, we take the horizontal domain  to be $[0.002, 0.02$], $[0.05, 0.2]$, and $[0.3, 3]$ for ${\rm I_{A,B}}$, ${\rm II_A}$ and ${\rm II_B}$, respectively. 

For understanding the influence of $g_X$, we show the situations of $g_X=(0.05, 1.0)$ for scheme $I_A$ and $g_{X}=(0.5, 1.0)$ for scheme ${\rm II_A}$. From the plots, it can be seen that when the value of $g_X s_\theta$ is fixed,  the contributions of $H^0$ are smaller in schemes ${\rm I_A}$ if the value of $g_X$ is larger. That is, $H^0$ in ${\rm I_A}$ has a significant effect at small $g_X$ or large $s_\theta$ values. 
Since $H^0$ is off-shell  in scheme ${\rm I_B}$ and its effect is negligible, by comparing the results of ${\rm I_A}$ with those of  ${\rm I_B}$, one can determine the influence of on-shell $H^0$ on ${\rm I_A}$.  
Since $h$ is an off-shell particle in scheme ${\rm II_A}$, the main contributions are from the invisible $H^0$ decays. According to Eq.~(\ref{eq:inv_H0}) and the results of Fig.~\ref{fig:Brhxx} and Table~\ref{tab:Nevent},  we need a somewhat larger value of $g_X s_\theta$ to get more signal events. Therefore,  the domain of $g_X s_\theta$ is set to be one order of magnitude larger than that in ${\rm I_{A,B}}$.  In scheme ${\rm II_B}$,  the signal events are from  off-shell $h$ and $H^0$. Therefore, in order to enhance the signal events, we need a large value of $g_X s_\theta$. Unfortunately, when $S>2$,   $g_X$ becomes a strong coupling constant. 
We thus omit the scheme ${\rm II_B}$ in the following discussions.      

\begin{figure}[tb]
\begin{center}
\includegraphics[width=70mm]{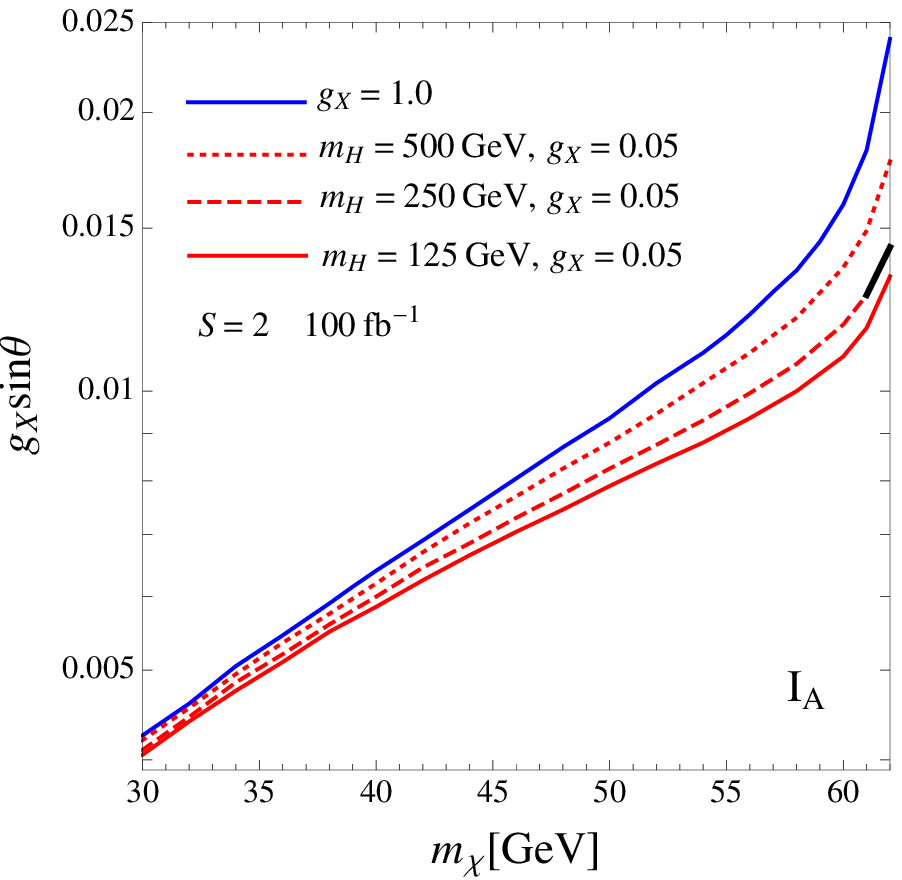}
\includegraphics[width=70mm]{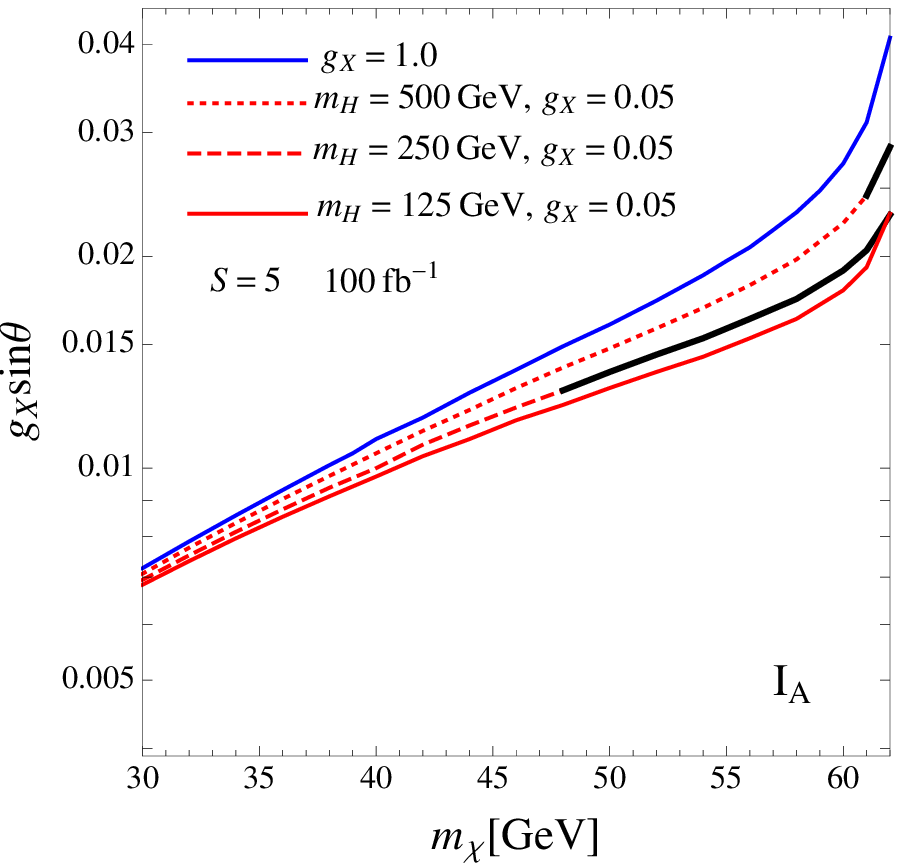}
\includegraphics[width=70mm]{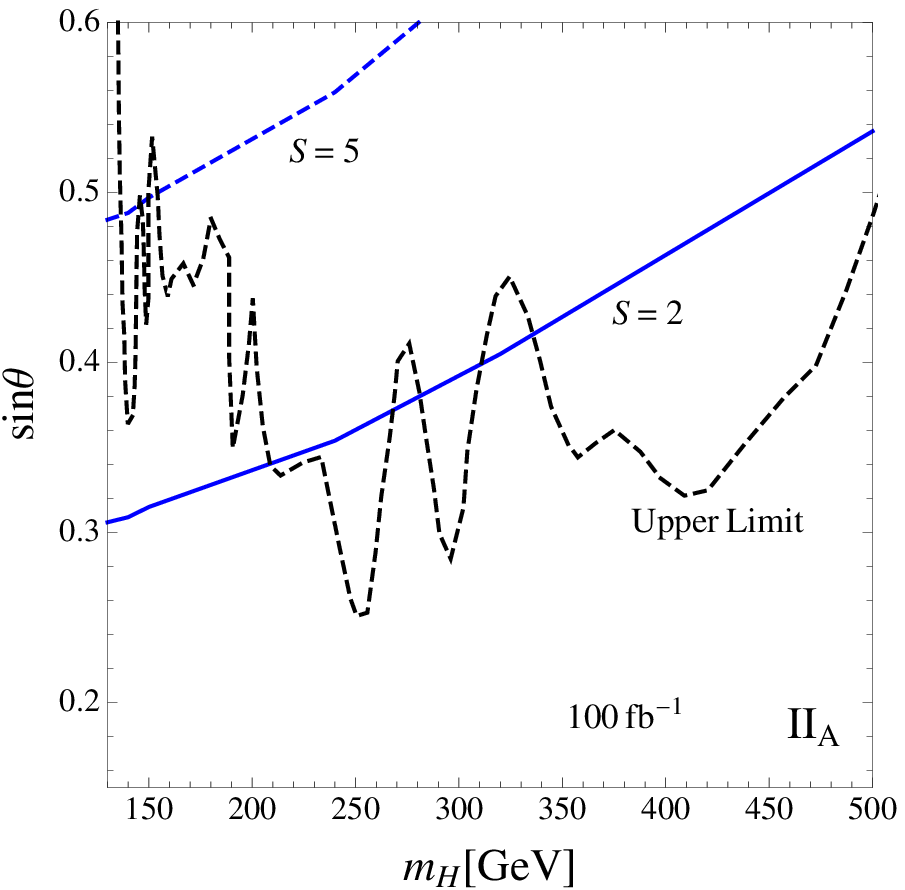}
\end{center}
\caption{  The upper left (right) plot shows the correlation between $g_X \sin \theta$ and $m_\chi$ for $S=2(5)$ in  scheme ${\rm I_A}$. The lower plot shows the correlation between $\sin \theta$ and $m_H$ in scheme ${\rm II_A}$. The upper limit of the mixing angle $\sin \theta$ is given from the analysis of the Higgs boson search at the LHC in Fig.~4 of Ref.~\cite{Robens:2015gla}. The region with thick black lines for the scheme ${\rm I_A}$ is excluded by the constraint on $\sin \theta$. }
\label{fig:mH_plot}
\end{figure}

Furthermore, in order to understand the  dependence of significance on the second Higgs mass, we investigate the significance by changing the value of $m_H$.  We note that scheme ${\rm I_B}$ is independent of $m_H$  as long as it satisfies $m_H < 2 m_\chi$  because off-shell $H^0$ effects are negligible. We thus focus on schemes ${\rm I_A}$ and ${\rm II_A}$ here.  The upper left (right) plot in Fig.~\ref{fig:mH_plot} shows the contours in $m_\chi$-$g_X \sin \theta$ plane with $S=2(5)$. For $g_X = 0.05$, we take $m_H = 125$, $250$ and $500$ GeV. Due to the small $\sin \theta$,   $H^0$ contributions are negligible and the $g_X =1$ case does not depend on $m_H$ . It can be seen that if the value of $g_X \sin \theta$ is fixed, the significance  increases with decreasing  $m_H$. Moreover the effect of $H^0$ can be seen for $2 m_\chi \sim m_h$, even though $m_H$ is as heavy as 500 GeV.  Also, the region with thick black lines is excluded from  the analysis of the Higgs boson search at the LHC in Fig.~4 of Ref.~\cite{Robens:2015gla}.  The lower plot in Fig.~\ref{fig:mH_plot} show the contours in the $m_H$-$\sin \theta$ plane with  $S=2$ and 5.  Since the branching ratio for $H^0 \to \chi \bar \chi$ is $\sim 1$, the significance in scheme ${\rm II_A}$ does not strongly depend on $g_X$  when  $g_X \gtrsim 0.1$. Moreover the significance does not depend on $m_\chi$ as long as $m_H > 2 m_\chi$ is satisfied. We also show the upper limit of $\sin \theta$, taken from Fig. 4 of Ref.~\cite{Robens:2015gla}, as a function of $m_H$.  We find that the stronger constraint for $\sin \theta$ is in the higher $m_H$ region.   

\begin{figure}[tb]
\begin{center}
\includegraphics[width=70mm]{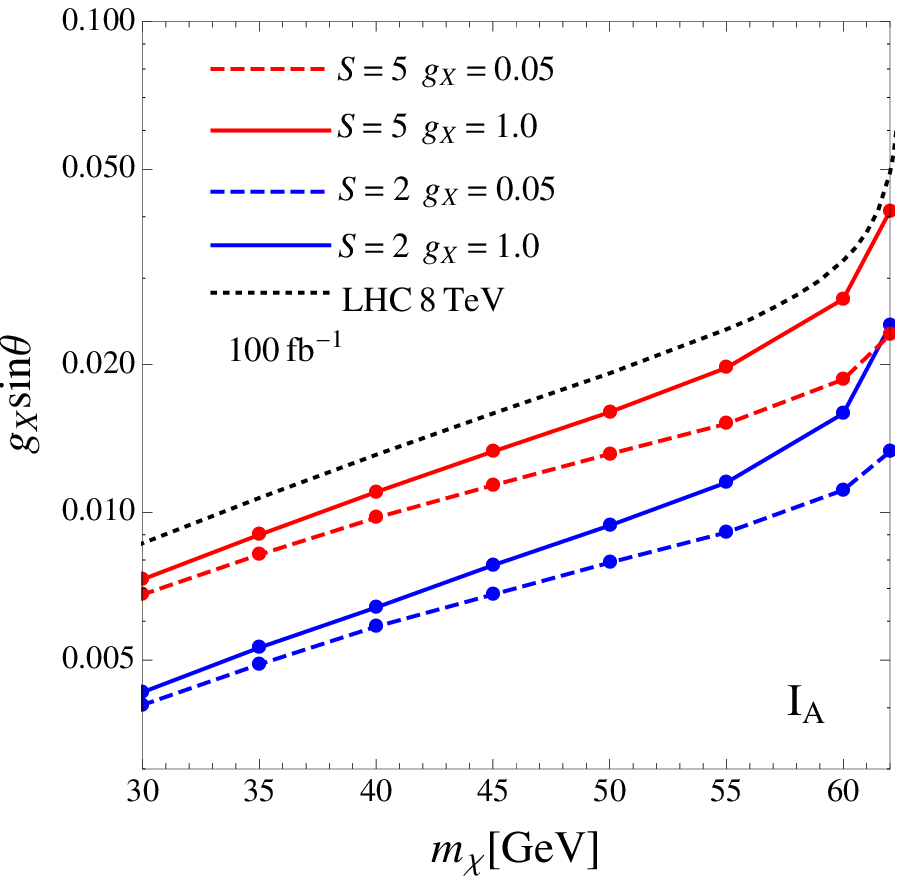}
\includegraphics[width=70mm]{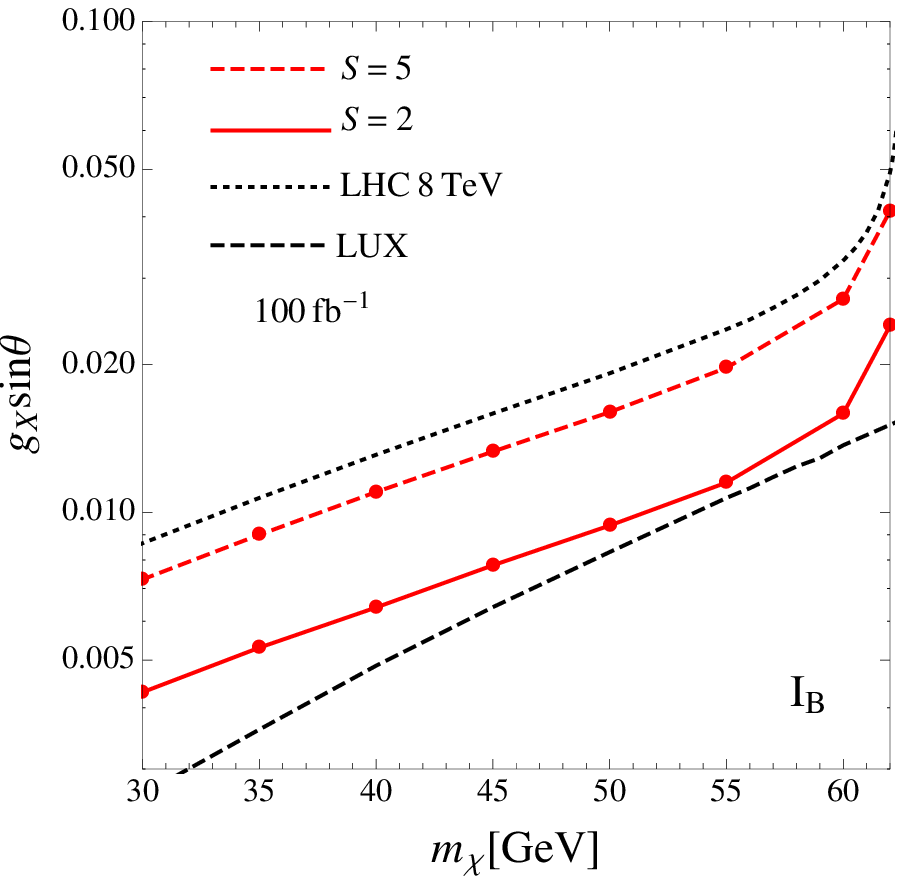}
\includegraphics[width=70mm]{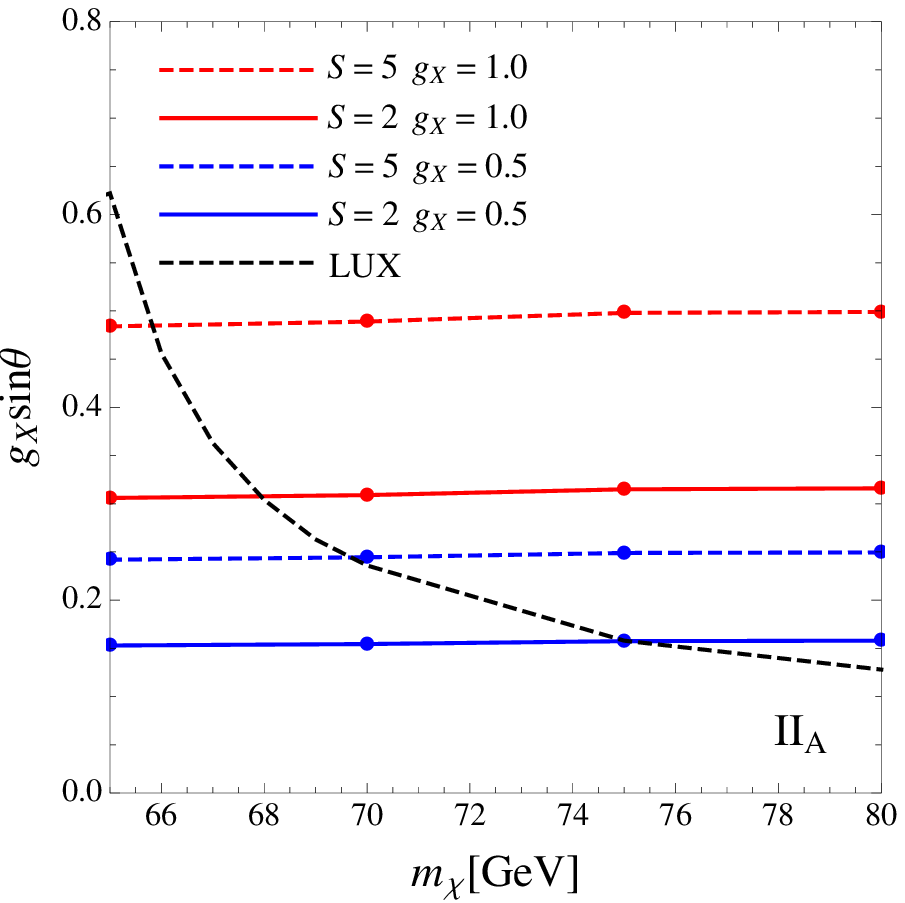}
\end{center}
\caption{ Correlation between $g_X \sin\theta$ and $m_\chi$ for  $S=2$ and $S=5$, where DM direct detection  measured by the LUX Collaboration~\cite{Akerib:2013tjd}  is included and the upper limit of invisible Higgs decay measured by ATLAS~\cite{Invisible_ATLAS} at $\sqrt{s}=8$ TeV is shown in schemes ${\rm I_{A,B}}$. }
\label{fig:Coupling_mX}
\end{figure}

After studying the potential for discovering DM signatures at the LHC,  in order to further understand the relationship between significance and free parameters, we show the correlation between $g_X s_\theta$ and $m_\chi$ for schemes ${\rm I_{A,B}}$ and ${\rm II_A}$ in Fig.~\ref{fig:Coupling_mX}.  
 Since the significance in the situation of lower $m_H$ and on-shell $H^0$ is larger, we also consider  $m_H = 2 m_\chi + 1$ GeV for schemes ${\rm I_A}$ and ${\rm II_A }$.
For illustration, we use 100 fb$^{-1}$ and adopt $S=2$ and $S=5$ in the plots. Since  the  large $g_X s_\theta$ in scheme ${\rm II_B}$ is excluded by DM direct detection experiments,  we do not further discuss the case.   
Additionally, the limit obtained from DM direct detection is also shown in the plots. 

 From the results in Fig.~\ref{fig:Coupling_mX},  it can be seen that the current invisible Higgs decay measured by ATLAS at $\sqrt{s}=8$ TeV~\cite{Invisible_ATLAS}  cannot give a strict bound on the parameters of schemes ${\rm I_{A}}$ and  ${\rm I_B}$. Due to off-shell $h$, the data of invisible Higgs decay  are not suitable for scheme ${\rm II_{A}}$. Recalling the results in Fig.~\ref{fig:Limit}, since the constraint from invisible Higgs decay in scheme ${\rm I_A ( I_B)}$ is stronger (weaker) than that from the LUX experiment, the significance over $S=2$ with 100 fb$^{-1}$ in scheme ${\rm I_B}$ is  excluded by the current LUX data. For enhancing the significance of scheme ${\rm I_B}$, a higher luminosity is necessary. Although the required values of $g_X s_\theta$ for $S=5$ in scheme ${\rm II_A}$ are one order of magnitude larger than those in scheme ${\rm I_B}$,  for the case with $m_{H^0} =2 m_\chi +1 > m_h$, $S=5$ is still allowed, even though the limit of the LUX is considered. We conclude that schemes ${\rm I_A}$ and ${\rm II_A}$ have the highest discovery potential in our model. 

\section{Conclusion}

A solution to the galactic center gamma-ray excess is DM annihilation  through the Higgs portal. We establish a Higgs portal model by considering a $SU(2)_X$ extension of the SM. We find that a $Z_3$ discrete symmetry is preserved when $SU(2)_X$ is broken spontaneously by the introduced quadruplet. Due to the residual $Z_3$ symmetry, the stable DM candidates in the model are the gauge bosons $\chi_\mu$ and $\bar\chi_\mu$. Besides the SM Higgs $h$, we have an extra scalar $H^0$ from the quadruplet. Since the quartic term $\Phi^\dagger_4 \Phi_4 H^\dagger H$ in the scalar potential leads to the mixture of  $h$ and $H^0$, the mixing angle $\theta$ plays an essential role in the connection between visible and invisible sectors and in DM relic density, DM direct detection, and gamma-ray excess \cite{Chen:2015nea}. 

In this paper, we studied the potential of  observing  invisible WIMPs at the 14-TeV LHC. Since VBF dominates the signal process, we only focused on this channel in the investigation. As a result, the signal events at the detector level are 2-jet + $\slashed{E}_T$. The possible backgrounds are from $Z/W$+ n-jet and $t W^- \bar b(\bar t W^+ b)$ with $n=2, 3$. In VBF, the DM pairs are produced by $h$ and $H^0$ portals, where $h$ and $H^0$ could be on-shell or off-shell, depending on the mass of DM. According to the mass of DM, we classify the interesting schemes to be $m_h /2 > m_\chi$, $m_h /2 < m_\chi < m_W$, $m_{H^0}  > 2m_\chi $, and $m_{H^0} < 2 m_\chi $, 
denoted as   ${\rm I_{A,B}}$ and ${\rm II_{A,B}}$, where ${\rm I (II)}$ stands for on-shell (off-shell) Higgs $h$ and  ${\rm A(B)}$ is on-shell (off-shell) $H^0$. 

We present the discovery significance of WIMPs with 100 and 300 fb$^{-1}$ at $\sqrt{s}=14$ TeV in Fig.~\ref{fig:Sig_Coupling}, where the bound from DM direct detection was not applied. From the plots, it can be seen that the four schemes used for numerical estimations could all reach a  significance of 5$\sigma$. 
However, in order to obtain a sizable significance, e.g., $S>2$, scheme ${\rm II_B}$ requires a strong coupling constant, which is excluded by the DM direct detection experiments.

Furthermore, in Fig.~\ref{fig:mH_plot}, we show the dependence of the significance on the $H^0$ mass  by concentrating on schemes ${\rm I_A}$ and ${\rm II_A}$, where the produced $H^0$ is  on-shell.  For ${\rm I_A}$, we find that the effect of $H^0$ would be seen for  $2 m_\chi \sim m_h$, even though $m_H$ is as heavy as 500 GeV. For ${\rm II_A}$, since the parameters are strongly constrained by the Higgs boson search, we find that in order to get $S > 2$ lighter $H^0$ is preferred for $\sin \theta$ to be large.

In Fig.~\ref{fig:Coupling_mX}, we show the correlation between $g_X \sin\theta$ and $m_X$ for $S=2$ and $S=5$ in ${\rm I_{A,B}}$ and ${\rm II_A}$, where the limit from LUX experiments is included. In the plots, we just use 100 fb$^{-1}$ as the representative value. From the results, we find that the values of parameters for $S=5$ in schemes ${\rm I_A}$ and ${\rm II_A}$ could satisfy the bound of the LUX experiments. Hence, the proposed DM scalar portal model could be tested by the data of  the 14-TeV LHC. \\

\noindent{\bf Acknowledgments}

 This work is supported by the Ministry of Science and Technology of 
R.O.C. under Grant \#: MOST-103-2112-M-006-004-MY3 (CHC) and MOST-103-2811-M-006-030 (TN). 

\end{document}